\documentstyle[aps,prc,preprint,epsfig,tighten]{revtex}
\begin{document}
\preprint{WU-HEP-99-3}
\draft
\title{Numerical analysis of two-pion correlation based on a hydrodynamical
model}
\author{Kenji Morita$^1$, Shin Muroya$^2$, Hiroki Nakamura$^1$, and Chiho
Nonaka$^3$}
\address{$^1$Department of Physics, Waseda University, Tokyo 169-8555, Japan\\
$^2$Tokuyama Women's College, Tokuyama, Yamaguchi 745-8511, Japan\\
$^3$Department of Physics, Hiroshima University, Higashi-hiroshima, Hiroshima
739-8526, Japan}
\date{\today}
\maketitle
\setlength{\unitlength}{1cm}
\begin{picture}(0,0)(0,0)
 \put(13.93,6.2){TWC-99-1}
\end{picture}

\begin{abstract}
 We will numerically investigate two-particle correlation function of
 CERN--SPS 158 A GeV Pb+Pb central collisions in detail based on a
 (3+1)-dimensional relativistic hydrodynamical model with first order phase
 transition. We use the Yano-Koonin-Podgoretski\u{\i} parametrization as well
 as the usual Cartesian parametrization and analyze the pair momentum
 dependence of HBT radii extracted from the parametrizations. We find that
 the interpretation of the temporal radius parameters as the time duration in
 YKP parametrization is not available for the hydrodynamical model where the
 source became opaque naturally because of expansion and surface dominant
 freeze-out.  Finally, effect of the phase transition on the source opacity
 is also discussed. 
\end{abstract}

\section{Introduction}
\label{sec:intro}

Ultra-relativistic heavy ion collision aimed at producing the Quark-Gluon  
Plasma (QGP) state is one of the most attracting problems in modern 
nuclear-particle physics \cite{QM97}.  In these decades, many kinds of
candidates for the signal of the production or existence of QGP are proposed. 
Two-particle correlation has been accepted as one of the probable 
observables for finding anomalous events in which QGP can be produced, 
because of the naive speculation that an extremely huge hot region will be 
produced if phase transition occurs \cite{Pratt_PRD33,Bertsch_PRC40}. In
quantum optics, estimating source size through quantum correlation of emitted
particles is a well-established method called the Hanbury-Brown Twiss(HBT)
effect \cite{HBT,GGLP}.

Usually, parameters obtained from the Gaussian fit of the correlation
functions are regarded as a radius of the particle source (HBT radius).
However, if the source is not static, the interpretation of the Gaussian
parameter is not straightforward \cite{Hama_Makhlin,Schlei_PLB293}.  In the
case of ultra-relativistic heavy ion collision, in addition to the highly
dynamical properties, many kinds of experimental limitations, such as
rapidity windows, make the situation more complicated.  In the recent
analyses, there exist two ways of parametrization in the Gaussian fit of the
two-particle correlation function; one is Cartesian parametrization
\cite{Chapman_HIP1,Chapman_PRL74}, and the other is
Yano-Koonin-Podgoretski\u{\i} parametrization
\cite{Yano_PLB78,Chapman_PRC52,Wu_EPJC1}. In both parametrizations, we can
obtain their respective  ``size" of particle sources. However different
parametrization gives different values and interpretation of their values is
not clear enough.

In this paper we evaluate two-particle correlation functions of pion based 
on a hydrodynamical model with phase transition. Parameters in the
hydrodynamical model are so tuned as to reproduce {\it single particle}
distributions of CERN 158 A GeV Pb+Pb collisions
\cite{Nonaka_TWC99}. Comparing the source size obtained through correlation
and freeze-out hypersurface of the hydrodynamical model, we investigate the
physical meaning of the HBT radii in both parametrizations. The opaque
property \cite{Heiselberg98} of the source is also analyzed based on a
hydrodynamical model. In a hydrodynamical model, particle emission takes place
on the freeze-out hypersurface which is thin but not an instant in
time. Expansion of the fluid also should naturally occur.  Hence, the
apparent opaque property of the freeze-out hypersurface is not trivial. The
influence of first order phase transition is also investigated.

In section \ref{sec:hydro}, we make a brief introduction to the hydrodynamical
model with phase transition.  In section \ref{sec:2picorrelation}, we
formulate two-particle correlation functions. Transverse momentum dependence
and rapidity dependence of source parameters are discussed in Section
\ref{sec:KTdep} and \ref{sec:Ydep}, respectively. Section
\ref{sec:Conclusion} is devoted to the concluding remarks.

\section{Hydrodynamical Model}
\label{sec:hydro}

Focusing our discussion to the central collisions, we may assume cylindrical
symmetry to the system and introduce cylindrical coordinates $\tau$, $\eta$,
$r$  and $\phi$ instead of the usual Cartesian coordinates $x$, $y$, $z$ and
$t$, where,
\begin{mathletters}
 \label{coordinate}
 \begin{eqnarray}
  t &=& \tau\cosh \eta,\\ \label{t}
  z &=& \tau\sinh \eta,\\ \label{z}
  x &=& r \cos \phi,\\  \label{x}
  y &=& r \sin \phi. \label{y}
 \end{eqnarray}
\end{mathletters} 

The four velocities of the fluid $U^{\mu}$ is given as,
\begin{mathletters}
\label{4velocity}
 \begin{eqnarray}
  U^{t}&=&U^{\tau}\cosh \eta +U^{\eta}\sinh \eta, \\ \label{ut}
  U^{z}&=&U^{\tau}\sinh \eta +U^{\eta}\cosh \eta, \\ \label{uz}
  U^{x}&=&U^{r}\cos \phi+U^{\phi}\sin \phi, \\ \label{ux}
  U^{y}&=&U^{r}\sin \phi-U^{\phi}\cos \phi. \label{uy}
 \end{eqnarray}
\end{mathletters}

Taking the constraint, $U^{\mu}U_{\mu}=1$, and cylindrical 
symmetry into account,
$U^{\mu}$ can be represented with two variables $Y_{\text{L}}$ and 
$Y_{\text{T}}$ \cite{Akase_PTP85},
\begin{mathletters}
\label{4velocity2}
 \begin{eqnarray}
 U^{\tau}&=&\cosh Y_{\text{T}} \cosh(Y_{\text{L}}-\eta), \\ \label{utau}
 U^{\eta}&=&\cosh Y_{\text{T}} \sinh(Y_{\text{L}}-\eta), \\ \label{ueta}
 U^{r}&=&\sinh Y_{\text{T}}, \\ \label{ur}
 U^{\phi}&=&0. \label{uphi}
 \end{eqnarray}
\end{mathletters}

As is well known, the relativistic hydrodynamical equation is 
given as
\begin{equation}
\partial_{\mu}T^{\mu\nu} = 0,
\end{equation}
and the energy momentum tensor of perfect fluid is given by
\begin{equation}
T^{\mu\nu} = EU^{\mu}U^{\nu}-P(g^{\mu\nu}-U^{\mu}U^{\nu})
\end{equation}
where $E$ and $P$ are energy density and pressure, respectively, and the 
metric tensor $g^{\mu \nu}$ is $g^{\mu \nu}$ = diag.$(1,-1,-1,-1)$.
In a hydrodynamical model, thermodynamical quantities such as $E$ and $P$ 
are treated as local quantities through 
temperature $T=T(x^{\mu})$.

In order to solve the above hydrodynamical equation, we need an equation of
state. In this paper, we adopt the bag model equation of state as the
simplest model for the first order QCD phase transition. In the bag model,
thermodynamical quantities below the phase transition temperature correspond
to a massive hadronic gas. On the other hand, in the high
temperature phase, thermodynamical quantities are given by a massless
QGP gas with bag constant $B$.  In this paper, we assume pions and kaons in
the hadronic phase and u-, d-, s-quarks and gluons in the QGP phase.  Putting
the phase transition temperature, $T_{\text{c}}=160$ MeV, into the condition
of pressure continuity, we can obtain the bag constant as $B= 412$
MeV/fm$^3$. 

At the phase transition temperature $T_{\text{c}}$, pressure
$P$ is continuous but other quantities such as energy density $E$ and 
entropy density $S$ have discontinuity as a function of temperature.
We parameterize  these quantities during the phase transition by using 
the volume fraction $\lambda$ \cite{Alam_PREP273,Hioki_PLB261}.
We assume that thermodynamical quantities at $T = T_{\text{c}}$ are given as
the functions of the space-time point through $\lambda(x^{\mu})$,
\begin{mathletters}
 \label{mixedphase}
 \begin{eqnarray}
 E(\lambda) &=& \lambda E_{\text{QGP}}(T_{\text{c}})
  +(1 - \lambda) E_{\text{HAD}}(T_{\text{c}}), \\ \label{energydensity}
 S(\lambda) &=& \lambda S_{\text{QGP}}(T_{\text{c}}) 
 + (1 - \lambda) S_{\text{HAD}}(T_{\text{c}}), \label{entropydensity}
 \end{eqnarray}
\end{mathletters}
where $ 0 \le \lambda(x^{\mu}) \le 1 $. 
This parametrization enables us to solve the hydrodynamical equation 
easily.

We consider that the hadronization process occurs 
at freeze-out temperature $T_{\text{f}}$ = 140 MeV.
In the evaluation of the one-particle distribution 
in a hydrodynamical model, the Cooper-Frye formula \cite{Cooper_PRD10} 
is often used, however, there exist several delicate matters on the 
treatment of freeze-out hypersurface based on this formula 
\cite{Sinyukov_ZPHYSC43,Grassi_PLB355}.
In this paper, we analyze not only one-particle distribution 
but also two-particle distribution, hence, we adopt 
another type of the simplest formula which enables us to extend two-particle
distributions straightforward,
\begin{equation}
 \frac{d^{3}\Delta N_{i}}{d {\bf p}^{3}}
  =  
  \frac{U_{\mu}p^{\mu}}{\sqrt{{\bf p}^{2} + m^{2}_{i}}}
  \frac{1}{\exp \left( \frac{U_{\nu}p^{\nu}}{T} \right) -1}
  \frac{U_{\lambda}d\sigma ^{\lambda}}{(2\pi)^{3}}, \label{dis}
\end{equation}
where $p^{0}= \sqrt{ {\bf p}^{2} + m_{i}^{2} }$ and 
the subscript $i$ stands for  pion or kaon.
Integrating (\ref{dis}) on the freeze-out hypersurface $\Sigma$,  
we obtain the rapidity distribution,
\begin{equation}
 \frac{dN_{i}}{dY}
  =
  \int_\Sigma \frac{U_{\mu}p^{\mu}}{\exp \left( \frac{U_{\nu}p^{\nu}}{T}
								   \right) -1}
   \frac{U_{\lambda}d\sigma^{\lambda}p_{ T}dp_{ T}d\phi}{(2\pi)^{3}},
\end{equation}
and the transverse mass distribution,
\begin{equation}
 \frac{1}{m_{T}} \frac{dN_{i}}{dm_{T}}
  =
  \int_\Sigma \frac{U_{\mu}p^{\mu}}
  {\exp \left( \frac{U_{\nu} p^{\nu}}{T} \right)-1}
  \frac{U_{\lambda}d\sigma^{\lambda}dYd\phi}{(2\pi)^{3}}.
\end{equation}
In a previous paper, we have already analyzed one-particle distribution of
hadrons by using results of our hydrodynamical model and compared it to 
the data at CERN SPS experiment \cite{Nonaka_TWC99}. The detailed structure of
space-time evolution and particle distribution is discussed in the Appendix.

\section{Two-Pion Correlation Function}
\label{sec:2picorrelation}

Two-particle correlation function is defined as
\begin{equation}
 C({\bf k_1},{\bf k_2})=\frac{W({\bf k_1},{\bf k_2})}
  {W({\bf k_1})W({\bf k_2})}\, ,\label{c2def}
\end{equation}
where $W({\bf k})$ and $W({\bf k_1},{\bf k_2})$ are one-particle
distribution $E_k~dN/d^3{\bf k}$ and two-particle
distribution $E_{k_1} E_{k_2} ~dN / d^3{\bf k_1}d^3{\bf k_2}$,
respectively. Assuming that the source is completely chaotic, for simplicity,
the two-particle correlation function for identical
bosons (i.e.,pions) is rewritten as \cite{Morita_PTPS129}
\begin{equation}
 C({\bf k_1},{\bf k_2}) = 1 + \frac{|I({\bf k_1},{\bf k_2})|^2}
  {W({\bf k_1}) W({\bf k_2})}, \label{c2}
\end{equation}
where 
\begin{equation}
 I({\bf k_1},{\bf k_2}) 
  = \int_\Sigma U_\mu d\sigma^\mu \, 
  e^{i(k_1-k_2)_\nu x^\nu}
  \sqrt{F({\bf k_1},x)F({\bf k_2},x)}\, , 
\end{equation}
and
\begin{equation}
 F({\bf k},x) 
  =\frac{1}{(2\pi)^3}\frac{U_\mu k^\mu}{E_k}
  \frac{1}{\exp(\frac{U_\nu k^\nu}{T_f})-1}\, .
\end{equation}

As usual, we introduce the average momentum
$K^\mu=\frac{1}{2}(k_1^\mu+k_2^\mu)$ and the relative momentum
$q^\mu=k_1^\mu-k_2^\mu$ of two particles.  Taking an approximation that
${\bf k_1}\simeq {\bf k_2}\simeq {\bf K}$ in Eq.(\ref{c2}),
the two-particle correlation function $C({\bf k_1},{\bf k_2})$ 
can be expressed as
\cite{Chapman_HIP1,Pratt_PRC42}
\begin{equation}
 C(q^\mu,K^\mu) \simeq 
  1+\frac{|\int d^4x \, e^{iq_\mu x^\mu}\,S(x^\mu,K^\mu)|^2}
  {|\int d^4x \, S(x^\mu.K^\mu) |^2}, \label{c2_approximation} 
\end{equation}
where $S(x^\mu,K^\mu)$ is the source function,
\begin{equation}
 S(x^\mu,K^\mu) = \int_\Sigma U_\mu(x') d\sigma^\mu(x') E_K F({\bf K},x') 
  \delta^4(x-x')\ .
\end{equation}
Furthermore, expanding $e^{iq_\mu x^\mu}$ in Eq.(\ref{c2_approximation}) with
$q_\mu x^\mu \ll 1$, we can obtain the simple expression
\begin{equation}
 C(q^\mu,K^\mu)-1 \approx 
  1+\langle q_\mu x^\mu \rangle^2 - \langle (q_\mu x^\mu)^2 
  \rangle ,  \label{c2_expanded}
\end{equation}
with $\langle A(x^\mu) \rangle$ being weighted average ,
\begin{equation}
 \langle A(x^\mu) \rangle \equiv \frac{\int d^4x \, A(x^\mu) S(x^\mu,K^\mu)}
  {\int d^4x \, S(x^\mu,K^\mu)} \label{average} \, .
\end{equation}
Consequently, we can recognize the two-particle correlation function as the
second order moment of $q_\mu x^\mu$ with the weight of $S(x^\mu,K^\mu)$. 
Because of the {\it on-shell} property of final particles, the relative
momentum $q_\mu$ has three independent components. Choosing the three
independent components of $q^\mu$ in an appropriate manner, we can obtain
meaningful information on the space-time structure of the particle source.


In the estimation of the source sizes from the 
two-particle correlation function (\ref{c2}), 
two types of Gaussian parametrizations\cite{Chapman_PRC52} are widely used.
One is the Cartesian parametrization,
including the outward-longitudinal cross term \cite{Chapman_PRL74}, and the
other is so called Yano-Koonin-Podgoretski\u{\i} parametrization 
\cite{Yano_PLB78,Wu_EPJC1}.
We take the z-axis in the direction of the collision and 
the x-axis parallel to the transverse component of the 
average momentum $K^{\mu}=(K^0,K_T,0,K_L)$, as usual.  
Then the z-axis, x-axis and y-axis correspond to  so-called
``longitudinal'', ``outward'' and ``sideward'' directions, respectively.

In the Cartesian parametrization, the three independent components
of the relative momentum $q_\mu$ are 
 $(q_x,q_y,q_z)=(q_{\text{out}},q_{\text{side}},q_{\text{long}})$. 
In this parametrization, the fitting function with 
the average momentum $K^\mu$ is given as
\begin{equation}
 C(q^\mu,K^\mu)
=1+\lambda\exp
 \left[-R_{\text{side}}^2(K^\mu)q_{\text{side}}^2
       -R_{\text{out}}^2(K^\mu)q_{\text{out}}^2 
           -R_{\text{long}}^2(K^\mu) q_{\text{long}}^2 
           -2R_{\text{ol}}^2(K^\mu)q_{\text{out}}q_{\text{long}}\right].
 \label{cartesian}
\end{equation}
There are four fitting parameters in Eq.(\ref{cartesian}), namely,
$R_{\text{side}}$, $R_{\text{out}}$, $R_{\text{long}}$ and $R_{\text{ol}}$,
which are so-called HBT radii. As for $\lambda$, we fix its value one because
we have assumed chaotic sources.

In this parametrization,  the meaning of these HBT radii, which is
given through Eq.(\ref{c2_expanded}), depends on the observer's longitudinal
frame. The coordinate system where $\beta_l=K_L/K^0=0$ is called
longitudinal co-moving system (LCMS), and on this system, the
interpretation becomes very simple \cite{Chapman_HIP1},
\begin{mathletters}
 \label{Cart_R}
 \begin{eqnarray}
 R_{\text{side}}^2 \quad &=& \quad \left\langle\tilde{y}^2 \right\rangle ,\\
 R_{\text{out}}^2 \quad &=& \quad \left\langle (\tilde{x} -\beta_\bot 
 \tilde{t})^2 \right\rangle ,\\
 R_{\text{long}}^2 \quad &=& \quad \left\langle\tilde{z}^2 \right\rangle ,\\
 R_{\text{ol}}^2  \quad &=& \quad \left\langle (\tilde{x}-\beta_\bot
 \tilde{t})\tilde{z}\right\rangle ,
 \end{eqnarray}
\end{mathletters}
where $\tilde{x}=x-\langle x \rangle$ and $\beta_\bot=K_T/K^0$. It is clearly
seen that these radii mix spatial and temporal components.
Therefore, from this parametrization it is not easy to extract accurately
physical information such as emission duration \cite{Wu_EPJC1}.

In the Yano-Koonin-Podgoretski\u{\i} (YKP) parametrization, the three
independent components of relative momentum are chosen as
$q_\bot=\sqrt{q_{\text{side}}^2+q_{\text{out}}^2}$, $q_l=q_{\text{long}}$ and 
$q^0=E_1-E_2$. The fitting function is given as \cite{Wu_EPJC1}
\begin{equation}
  C(q^\mu,K^\mu)
 =1+\lambda\exp \left\{ -R_\bot^2(K^\mu)q_\bot^2
                                 -R_\|^2(K^\mu)\left[q_l^2-(q^0)^2\right]
                                 -\left[R_0^2(K^\mu)+R_\|^2(K^\mu)\right]
                                 [q_\mu u^\mu(K^\mu)]^2\right\} ,\label{ykp}
\end{equation}
where
$u(K^\mu)=\gamma(K^\mu)(1, 0, 0, v(K^\mu))$, and
$\gamma(K^\mu)=1/\sqrt{1-v^2(K^\mu)}$. We fixed $\lambda=1$ as well as in the
Cartesian parametrization. Hence, the fitting parameters are $R_\bot$, $R_0$,
$R_\|$ and the so-called Yano-Koonin-Podgoretski\u{\i} velocity $v$ which
appears in the four-velocity $u^\mu$. The first three parameters are the HBT
radii that are invariant under longitudinal boosts. This is one of the most
important properties of the YKP parametrization. The four-velocity $u^\mu$ has
a longitudinal component only, and the Yano-Koonin-Podgoretski\u{\i} velocity
$v$ can be interpreted as the longitudinal source velocity \cite{Wu_EPJC1}.
The interpretation of the HBT radii in the YKP parametrization can be given in
the same manner as in Cartesian parametrization.  Denoting deviation as
$(\Delta A)^2= \left \langle (A-\langle A \rangle)^2 \right \rangle$ and
\begin{mathletters}
 \label{YKP_ABC}
 \begin{eqnarray}
 A&=&(\Delta t)^2 - \frac{2}{\beta_\bot} \langle \tilde{x} \tilde{t}
 \rangle + \frac{1}{\beta_\bot^2}
 \left[(\Delta x)^2-(\Delta y)^2 \right]\label{ykpA} ,\\
 B&=&(\Delta z)^2 - \frac{2\beta_\|}{\beta_\bot} \langle \tilde{x}
 \tilde{z} \rangle + \frac{\beta_\|^2}{\beta_\bot^2}
 \left[(\Delta x)^2-(\Delta y)^2\right]\label{ykpB} ,\\
 C&=& \langle \tilde{z} \tilde{t} \rangle -
 \frac{\beta_\|}{\beta_\bot} \langle \tilde{t}\tilde{x} \rangle
 -\frac{1}{\beta_\bot} \langle \tilde{z} \tilde{x} \rangle +
 \frac{\beta_\|}{\beta_\bot^2} \left[(\Delta x)^2-(\Delta y)^2\right],
 \label{ykpC}
 \end{eqnarray}
\end{mathletters}
the HBT radii and Yano-Koonin-Podgoretski\u{\i} velocity are expressed as
follows \cite{Wu_EPJC1} :
\begin{mathletters}
 \label{YKP_R}
 \begin{eqnarray}
 v(K^\mu)&=&\frac{A+B}{2C}
  \left[1-\sqrt{1-\left(\frac{2C}{A+B}\right)^2}\right]\label{ykpvelo}\\
 R_\bot^2(K^\mu)&=&(\Delta y)^2\label{ykprt}\\
 R_0^2(K^\mu)&=&A-vC \label{ykpr0}\\
 R_\|^2(K^\mu)&=&B-vC .\label{ykprl}
 \end{eqnarray}
\end{mathletters}
Even in the YKP parametrization, $R_\bot$ directly corresponds to transverse
size. In the Yano-Koonin-Podgoretski\u{\i}(YKP) frame which is defined by
$v=0$, these expressions (\ref{ykprt}-\ref{ykprl}) become very simple,
\begin{mathletters}
 \label{inYKPframe}
 \begin{eqnarray}
 R_\bot^2(K^\mu)&=&(\Delta y)^2 \label{rt} \ ,\\
 R_0^2(K^\mu)&=&(\Delta t)^2 - \frac{2}{\beta_\bot} \langle
 \tilde{x} \tilde{t} \rangle + \frac{1}{\beta_\bot^2}
 \left[(\Delta x)^2-(\Delta y)^2 \right]\label{r0}\ , \\
 R_\|^2(K^\mu)&=&(\Delta z)^2 - \frac{2\beta_\|}{\beta_\bot} \langle
 \tilde{x} \tilde{z} \rangle + \frac{\beta_\|^2}{\beta_\bot^2}
 \left[(\Delta x)^2-(\Delta y)^2\right]\label{rl}.
 \end{eqnarray}
\end{mathletters}
The values of these quantities do not change because they are invariant under
longitudinal boosts. But now the $\langle \cdots \rangle$ is taken in the YKP
frame. It has been shown that the last two terms in Eqs.(\ref{r0}) and
(\ref{rl}) can be considered as a small correction within some class of the
thermal model \cite{Wu_EPJC1}. Hence, the three HBT radii can be considered to
measure the ``source size'' directly. Specifically, $R_0$ measures the emission
duration on which, according to a naive expectation, the effect of first order
phase transition may be observed. Of course, it is not clear that this
interpretation works well even in the hydrodynamical model. 

\section{Transverse Momentum Dependence of Source Parameters}
\label{sec:KTdep}

In this section, we will discuss the dependence of HBT radii on the
transverse momentum $K_T$.  Because of the lack of experimental data with
fixed average momenta, in order to make a comparison with the experimental
data, we integrate the distribution functions in (\ref{c2}) with respect to
average momenta and
\begin{mathletters}
 \begin{eqnarray}
  C(q^\mu ; K_T)
   &=& 1+\frac{\int dK_L \,|I({\bf k_1},{\bf k_2})|^2}
   {\int dK_L\,W({\bf k_1}) W({\bf k_2})} , \label{c2_kint}\\
  C(q^\mu ; Y)
   &=& 1+\frac{\int K_T dK_T \,|I({\bf k_1},{\bf k_2})|^2}
   {\int K_T dK_T\,W({\bf k_1}) W({\bf k_2})} , \label{c2_yint}
 \end{eqnarray}
\end{mathletters}
is used.

\begin{figure}
 \begin{center}
  \epsfig{file=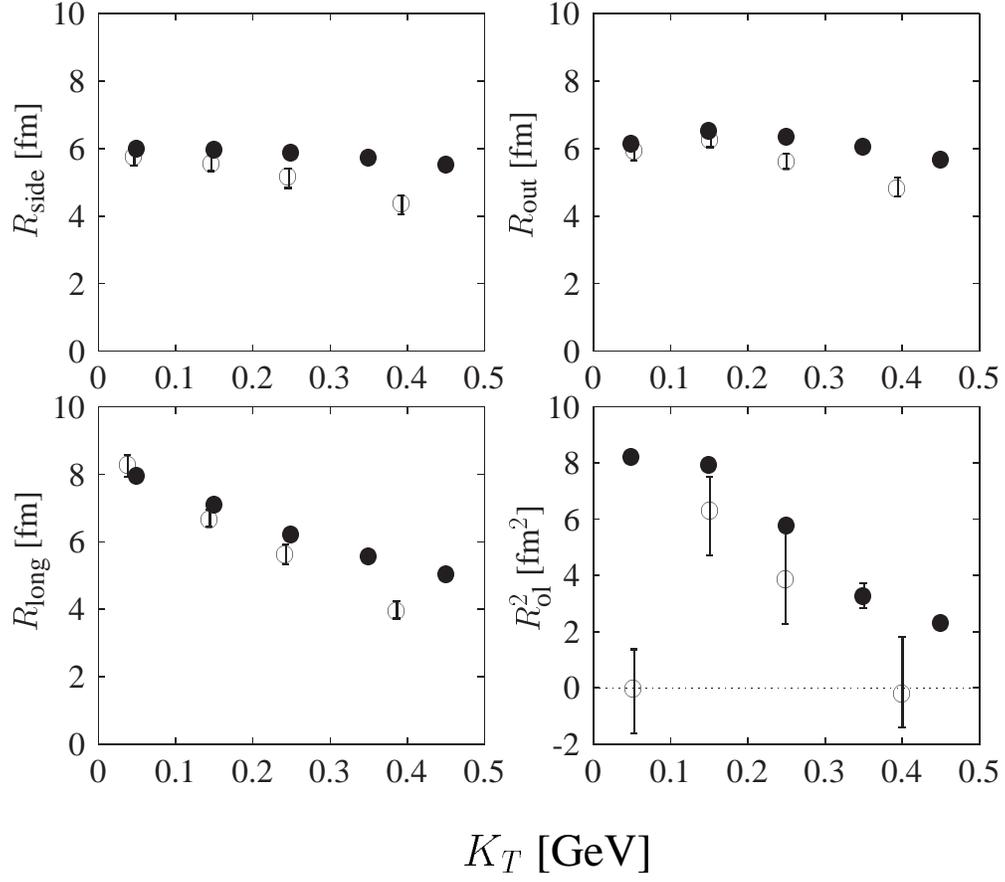}
 \end{center}
 \caption{$K_T$ dependence of Cartesian HBT radii at the rapidity of 
 $3 \leq Y_{\pi\pi} \leq 4$. The reference frame is at $Y_0=3.5$. Open circles
 show the experimental results taken from NA49 preliminary data
 \protect\cite{NA49_QM96}. Closed circles show our results extracted from
 Eq.\ (\protect\ref{cartesian}) through the Gaussian fitting.}
 \label{cartfig} 
\end{figure}

Figure \ref{cartfig} shows the dependence of Cartesian HBT radii on the average
transverse momentum $K_T$. Closed circles stand for our numerical 
results obtained from the correlation function (\ref{c2_kint}) through the
fitting function (\ref{cartesian}). Experimental results
taken from NA49 preliminary data \cite{NA49_QM96} are plotted as open
circles. Our results almost agree with the experimental data. However, we get
the HBT radii a little larger than experimental results at high $K_T$. 
In the upper left figure of Fig. \ref{cartfig}, though experimental
$R_{\text{side}}$ becomes smaller at larger $K_T$, this tendency appears only
weakly in our simulation. This $K_T$ dependence of $R_{\text{side}}$ can be
explained by the existence of strong transverse flow \cite{Schlei_PLB293} and
contributions from resonance decay \cite{Schlei_PRC54}. Hence, the
differences between our calculation and experimental results might be caused
by the simplification that neglected the resonance decay. 

\begin{figure}[ht]
 \begin{center}
  \epsfig{file=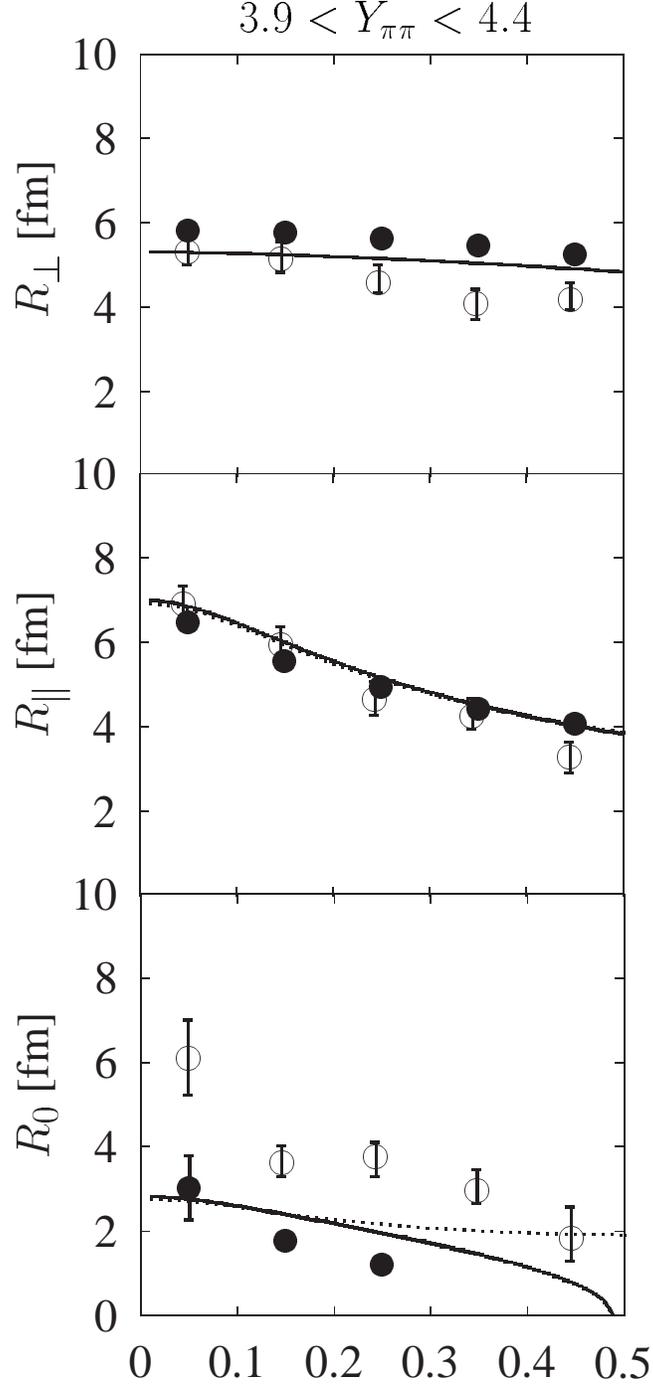}
 \end{center}
  \caption{$K_T$ dependence of YKP radii. Being the same as in Fig.\
 \protect\ref{cartfig}, open and closed circles show the experimental data
 and HBT radii, respectively. Solid lines stand for space-time extensions
 (\protect\ref{rt})-(\protect\ref{rl}). Dotted lines stand for source
 sizes. ($\Delta z$ for $R_\|$ and  $\Delta t$ for $R_0$.)}
  \label{ykpkt}
\end{figure}

As we discussed in the previous section, a simple interpretation is not 
available to HBT radii in the Cartesian parametrization; investigating the
source parameters based on the YKP parametrization is important.
%
%
Figure \ref{ykpkt} shows the dependence of YKP radii on average transverse
momentum $K_T$. As mentioned in Sec.\ \ref{sec:2picorrelation}, 
in the YKP parametrization, 
the HBT radii are expected to stand for longitudinal invariant 
``source size''. In Fig.\ \ref{ykpkt}, closed circles stand for YKP radii
obtained from the correlation function (\ref{c2_kint}) through fitting based
on Eq.(\ref{ykp}). Solid lines stand for the space-time extensions of 
freeze-out hypersurface evaluated from the right-hand side of
Eq.(\ref{rt})-(\ref{rl}). 
These quantities are expected to agree with our HBT radii
because in our calculation we consider only thermal pions
\cite{Schlei_PRC55}. Dotted lines shows ``source sizes''
($\Delta z$ for $R_\|$ and $\Delta t$ for $R_0$) calculated from the second
order moment, {\it i.e.}, first terms only in the right-hand side of
Eq.(\ref{r0})-(\ref{rl}).  In the calculation of space-time extensions and
``source sizes'', average momenta are fixed at the central value of the
integration range. NA49 experimental data \cite{NA49_EPJC2} are plotted as
open circles.

The upper figure shows the dependence of the transverse source radii $R_\bot$
on $K_T$. As we have already discussed,  $R_\bot$ in the YKP
parametrization is the same as $R_{\text{side}}$ in the Cartesian
parametrization. The difference between the space-time extension and YKP
radius $R_\bot$ is about 0.5fm.

The middle part of Fig.\ \ref{ykpkt} shows the dependence of longitudinal
source radius $R_\|$ on $K_T$. Our result from the fit is consistent with
both the experiment and space-time extension. Furthermore, the ``source
size'' $\Delta z$ also agrees with the space-time extension. Hence, the usual 
interpretation of $R_\|$ in the YKP parametrization works well in our
calculation.

The lower figure shows the dependence of the temporal source parameter $R_0$
on $K_T$. Our results from the fit are a few fm smaller than experimental
results and become smaller as $K_T$ increases. The space-time
extension calculated from Eq.(\ref{r0}) shows similar behavior in $K_T$ but
quantitatively agreement is not so good at large $K_T$. Because of the
on-shell condition, the available $q^\mu$ region is limited and a great
ambiguity in fitting process can occur. For example, $R_0^2$ sometimes tends to become
negative if we take the most likely values without any restriction to
$R_0^2$.  However, it becomes small but positive definite values if we
impose the restriction on $R_0^2$ to be positive. As for the ``time
duration'' $\Delta t$, though it agrees with $R_0$ at low $K_T$, it becomes
larger than $R_0$ at large $K_T$. Hence, the interpretation that $R_0$
measures the time duration directly in the YKP parametrization is not correct
in this region. This can be explained by the effect of ``source opacity''
\cite{Heiselberg98}, which will be discussed later. On the other hand, the
values of $R_0$ and $\Delta t$ are about 2-3 fm, obviously smaller than
$R_\bot$, $R_\|$. Hence, we can say that clear evidence of the first order
phase transition does not appear in the time duration. It was already pointed
out that the large time duration (or longer lifetime) might not occur for
Bjorken-type scaling, the  initial condition we used
\cite{Pratt_PRD33,Hung_PRL75}.

\begin{figure}[ht]
 \begin{center}
  \epsfig{file=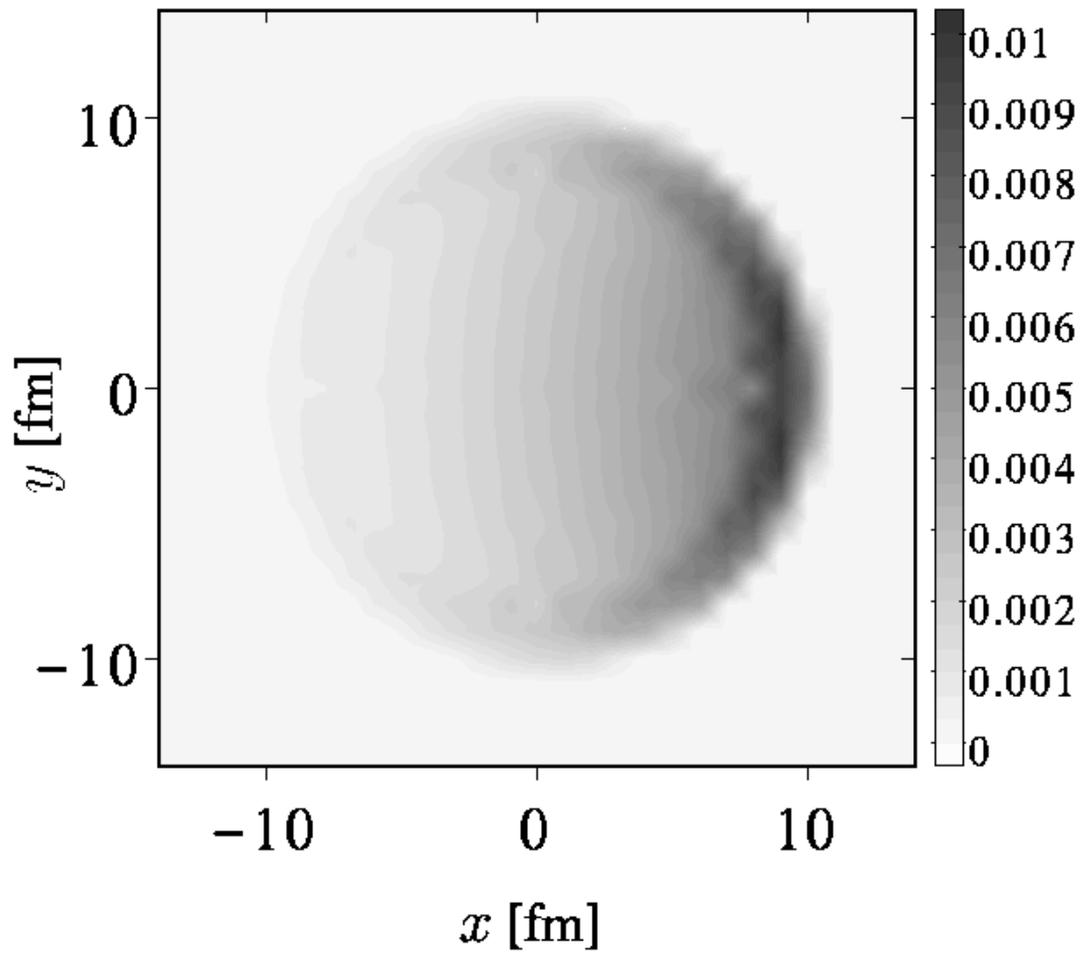}
 \end{center}
 \caption{Source function $\bar{S}_T(x,y)$. Thick regions emit particles more
 than thin regions.}
 \label{opaque1}
\end{figure}

The ``opaque'' source emits particles dominantly from the surface and the 
``transparent'' source emits from whole region. Whether the ``real'' source is
opaque or transparent is an interesting problem \cite{Heinz_QM97}. Aiming for
this point, we analyze the source function $S(x^\mu,K^\mu)$ for $K_T=450$ MeV
and $Y_{\pi\pi}=4.15$. Figure\ \ref{opaque1} shows the source function as a
function of transverse coordinates $x$ and $y$. Here the source function is
integrated in the temporal and longitudinal coordinates, $\bar{S}_T(x,y)=*\int dz \,
dt \, S(x^\mu,K^\mu)$, and normalized on the transverse plane as 
$\int dx \, dy \, \bar{S}_T(x,y)=1$. In this figure, emitted pions have only $x$
component of momentum. Figure\ \ref{opaque1} clearly shows that pions are
emitted mostly from the crescent surface region and the source is thinner in
the $x$-direction than in the $y$-direction. This is the typical property of
the opaque source. Therefore, the difference between $\Delta x$ and $\Delta y$,
that is, the difference of space-time extensions $(\Delta x)^2-(\Delta y)^2$
is a good measure of the opaque property\cite{Tomasik_TPR98}.

\begin{figure}[ht]
 \begin{center}
  \epsfig{file=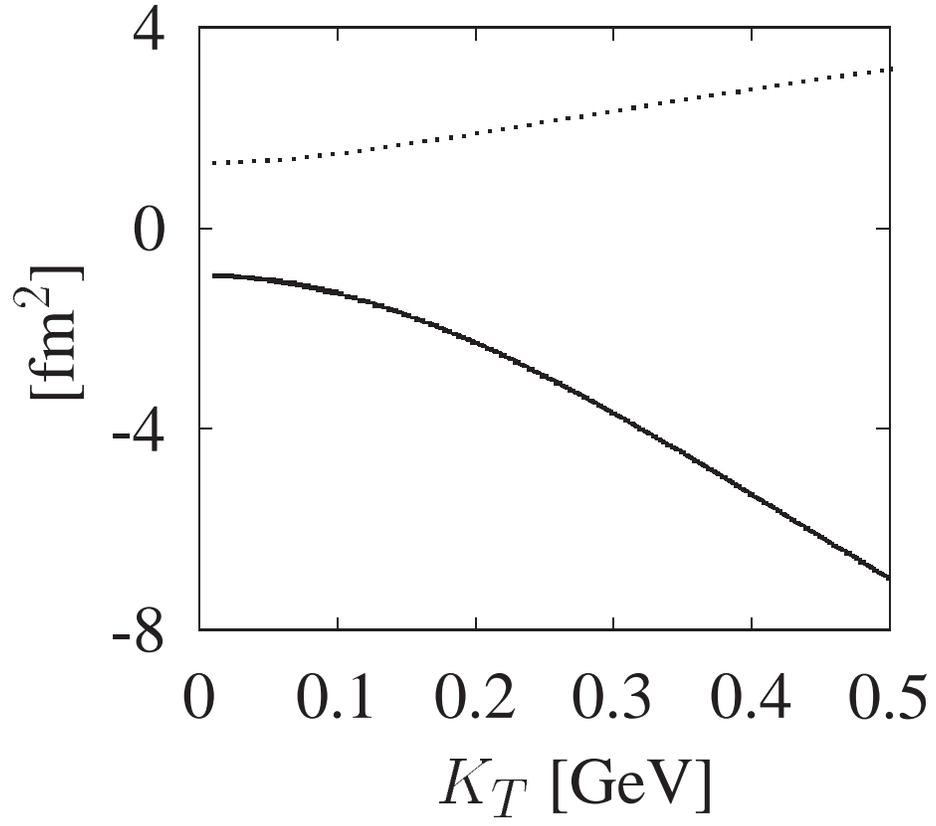}
 \end{center}
 \caption{The second term $-2\langle \tilde{x}\tilde{t} \rangle /\beta_\bot$
 (dotted line) and third term $((\Delta x)^2-(\Delta y)^2)/\beta_\bot^2$
 (solid line) in the right-hand side of Eq.(\protect\ref{r0}) are functions
 of $K_T$. $Y_{\pi\pi}$ is fixed at 4.15.}
 \label{opacity}
\end{figure}

Figure \ref{opacity} shows the dependence of the second and third terms
in Eq.(\ref{r0}) on $K_T$.  The third term $((\Delta x)^2-(\Delta
y)^2)/\beta_\bot^2$ has a large negative contribution to $R_0$ at a large $K_T$
region, while the contribution from the second term $-2 \langle \tilde{x}
\tilde{t} \rangle /\beta_\bot$ is not so large. Hence we may not consider the
third term as a small correction to $(\Delta t)^2$ and this is the reason why
the simple interpretation of $R_0$ in the YKP parametrization is not
available in our calculation. A simple model analysis \cite{Tomasik_TPR98}
has shown the contradictory result that $R_0$ diverges to $-\infty$ as
$K_T\rightarrow 0$ ($\beta_\bot \rightarrow 0$) for the opaque
source. This is due to the tendency of the denominator of the 
third term in Eq.(\ref{r0}) to go to zero as $K_T\rightarrow 0$  while the
numerator is kept as finite. However, this divergence does not occur in our
model because the numerator, $(\Delta x)^2-(\Delta y)^2$, goes to 0 as $K_T
\rightarrow 0$. 

\begin{figure}[ht]
 \begin{center}
  \epsfig{file=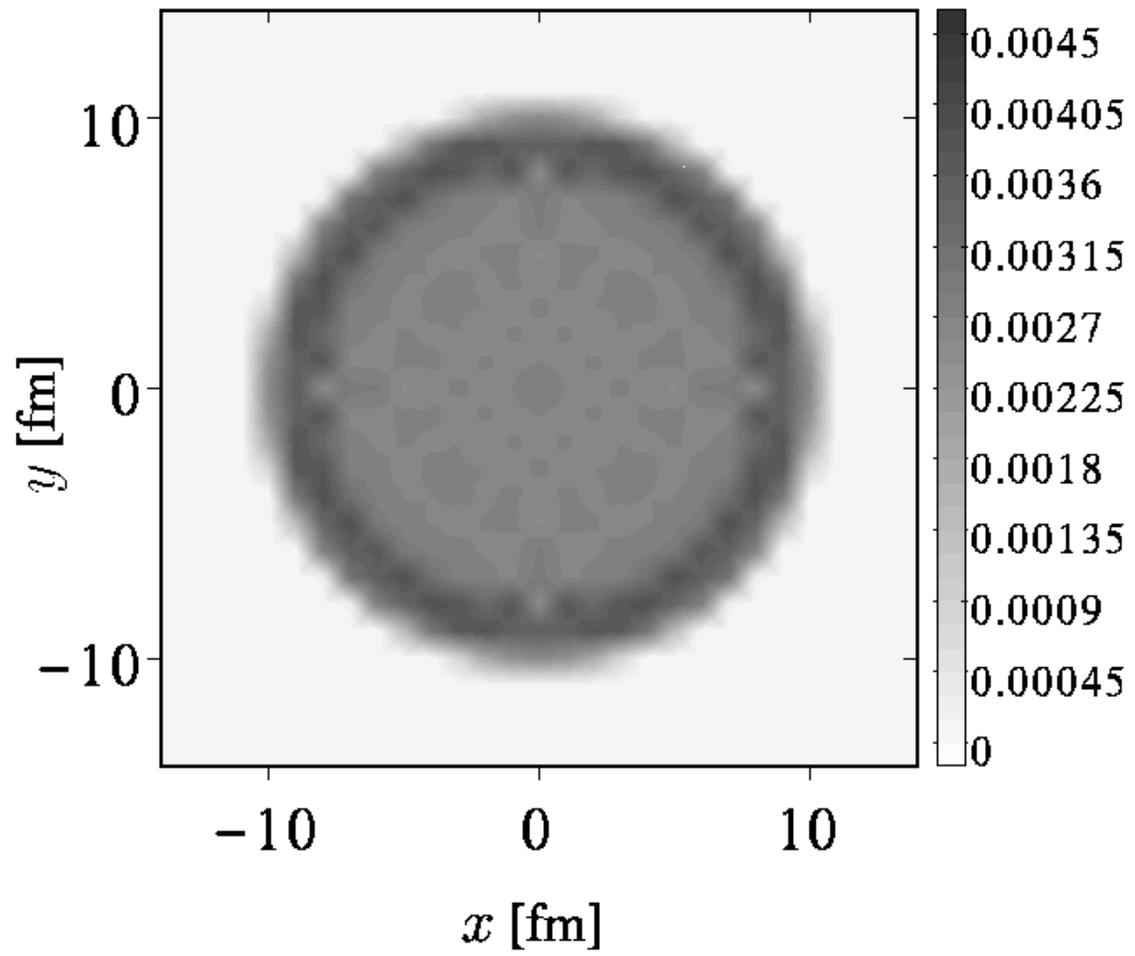}
 \end{center}
 \caption{Same as Fig.\protect\ref{opaque1}, but now the effect of transverse
 flow is neglected.}
 \label{opaque2}
\end{figure}

Figure \ref{opaque2} also shows the source function $\bar{S}_T(x,y)$ with the 
artificially neglected transverse flow. In order to see the effect of
transverse flow on the source function, here we put the transverse flow
velocity $U^r=0$ by hand. In this case, the source function $\bar{S}_T(x,y)$ is
almost proportional to the space-time volume of freeze-out hypersurface
projected onto the transverse plane, and the azimuthal symmetry of the source
function is restored as shown in Fig.\ \ref{opaque2}. As a result, the
measure of the source opacity, $(\Delta x)^2-(\Delta y)^2$, vanishes. 
Although particle emission from the inside is less than that from the
surface; the source is not opaque in this sense. When the transverse flow
exists, the source function is deformed by the thermal Boltzmann factor
$\exp(\cosh Y_T \cos \phi \, K_T /T_f)$ and in our simulation the transverse
flow rapidity, $Y_T$, is almost proportional to $r$ as shown in Fig.\
\ref{yt}. Consequently, the number of emitted particles in $x$-direction is
increased for the positive $x$ region ($\cos \phi > 0$), especially at
the surface, because of the behavior of $Y_T$, and the number is decreased
for the negative $x$ region ($\cos \phi <0$). This flow effect deforms
surface dominant distribution in Fig.\ \ref{opaque2} to the crescent shape.
This is the main reason why the source is opaque in our model. Though in the
recent experimental result, $R_0$ is larger than expected for the opaque
source. According to the scenario as discussed in Ref.\cite{Hung_PRL75}, the
phase transition may cause a large time duration for the opaque source.
However, in our hydrodynamical model, we assumed the first order phase
transition, but the time duration $\Delta t$ does not become large enough to
reproduce the experiment.

\begin{figure}[ht]
 \begin{center}
  \epsfig{file=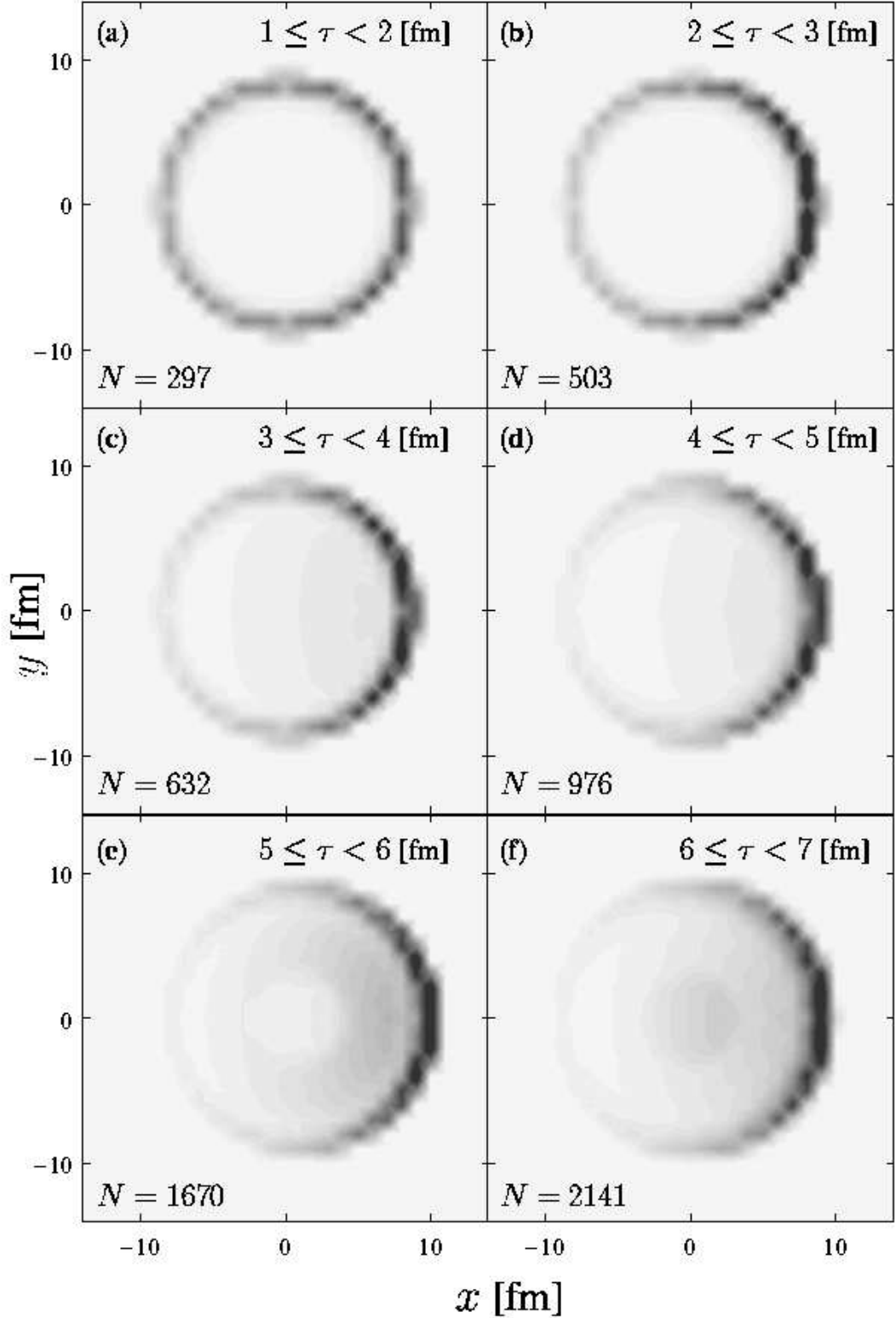}
 \end{center}
 \caption{Source functions $S_T(x,y)$ sliced by freeze-out proper
 time. The thickness in each figure represents the intensity of the source.}
 \label{taufig1}
\end{figure}
\begin{figure}[ht]
 \begin{center}
  \epsfig{file=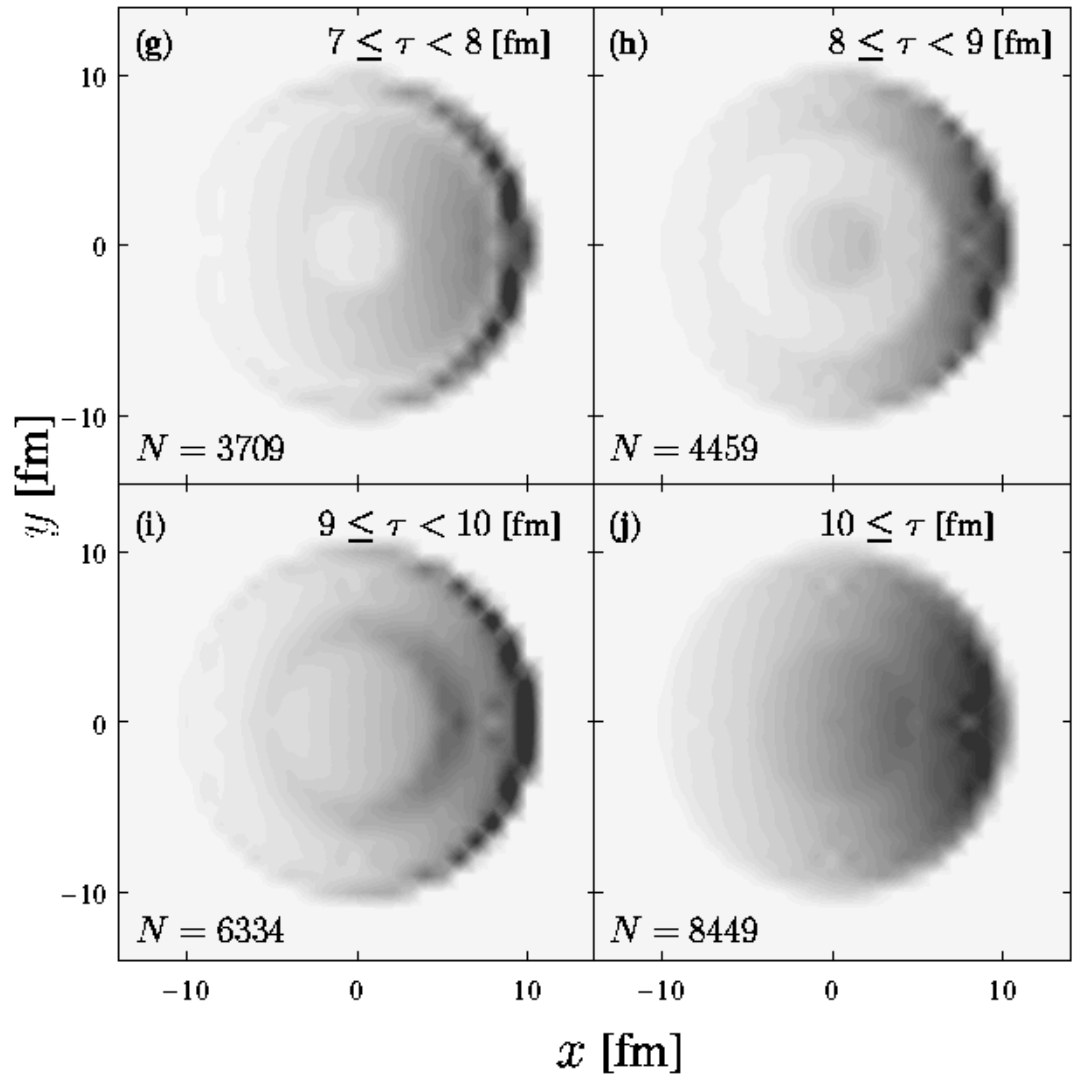}
 \end{center}
  \caption{Rest of Fig.\ \protect\ref{taufig1}}
  \label{taufig2}
\end{figure}

In order to discuss the effects of the phase transition on the source
opacity,  we sliced the source function $S_T(x,y)$ by the freeze-out proper
time $\tau$ and show the source functions for each proper time in Fig.\
\ref{taufig1},\ref{taufig2}. Figures\ \ref{taufig1}(a)-\ref{taufig2}(j) show
the same quantity as Fig.\ \ref{opaque1}, but they are not
normalized. Integral $N=\int dx \, dy \,S_T(x,y)$ should be considered as the
number of particles. Figures\ \ref{taufig1}(a)-\ref{taufig1}(f) clearly show
most particles are emitted from a thin surface. The measure of the source
opacity $(\Delta x)^2-(\Delta y)^2$ evaluated for Figs.\
\ref{taufig1}(b)-\ref{taufig2}(h) indicates extremely opaque property. For
these sources, the YKP parametrization cannot be used because the
YKP velocity becomes imaginary ({\it i.e.}, $A+B < 2C$ in Eq.(\ref{ykpvelo}))
\cite{Tomasik_TPR98}. Figure\ \ref{taufig1}(a) shows strong surface
emission, but this source does not show the source opacity because the
transverse flow is very weak at the early stage of evolution. $(\Delta
x)^2-(\Delta y)^2$ is negative throughout Fig.\ \ref{taufig1},\ref{taufig2}. 
As discussed in the Appendix, if first order phase transition
exists, a flat-top structure appears in temperature distribution (Fig.\
\ref{temperature}) which corresponds to a large region caused by latent heat
released at phase transition. This equi-temperature region reaches freeze-out
temperature almost simultaneously and makes a filled source function Fig.\
\ref{taufig2}(j). Hence, the existence of phase transition weakens the source
opacity.

\section{Rapidity Dependence of Source Parameters}
\label{sec:Ydep}

\begin{figure}[ht]
 \begin{center}
  \epsfig{file=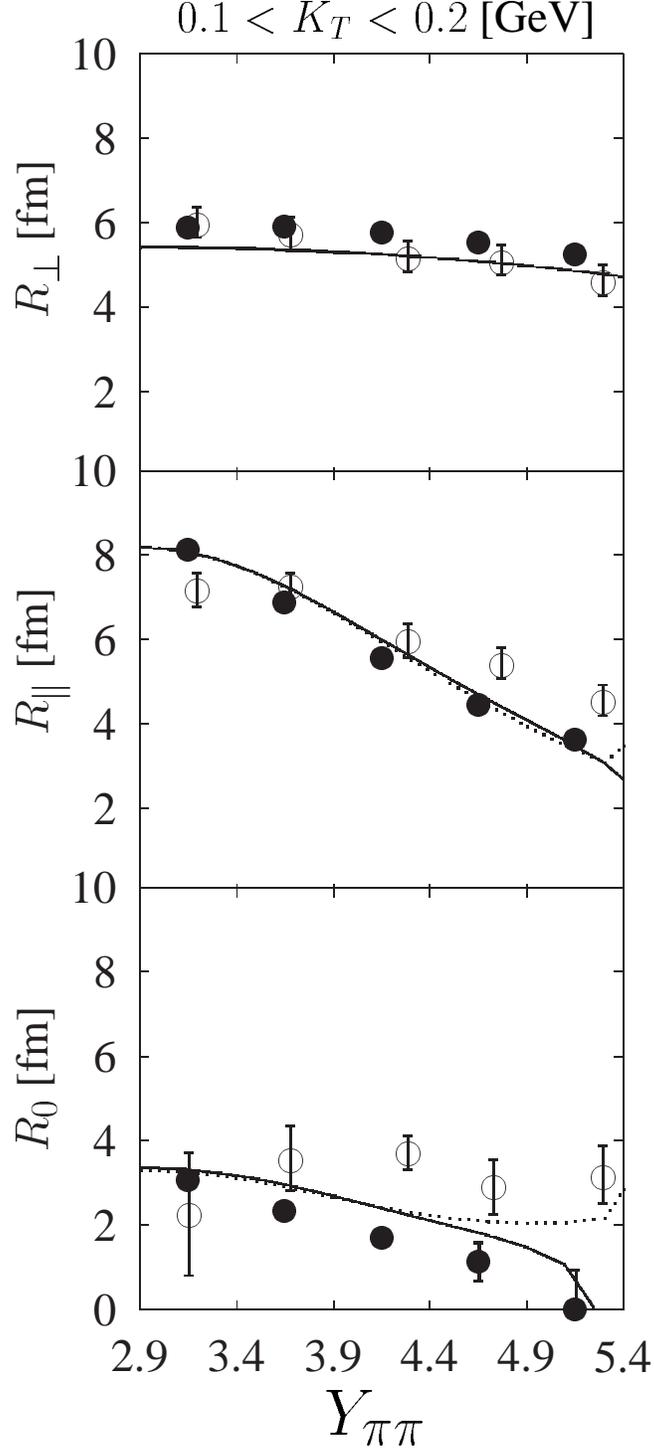}
 \end{center}
 \caption{Pair rapidity $Y_{\pi\pi}$ dependence of YKP radii. As in
 Fig.\ \protect\ref{ykpkt}, open circles, closed circles, solid lines and
 dotted lines stand for experimental results, our HBT radii, space-time
 extensions and source sizes, respectively.}
 \label{ykpy}
\end{figure}

\begin{figure}[ht]
 \begin{center}
  \epsfig{file=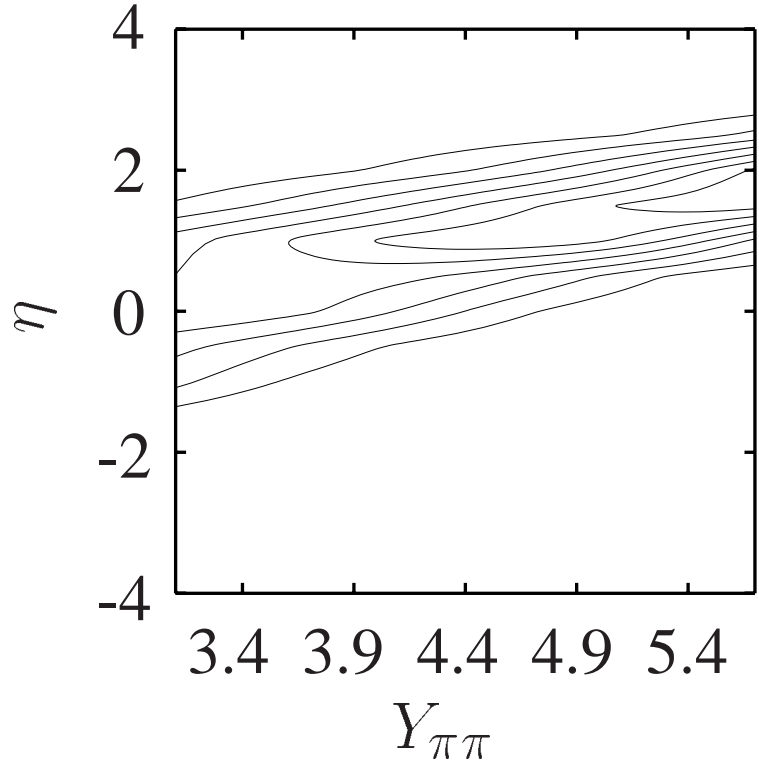}
 \end{center}
 \caption{Contour plot of source function for our model as a function of
 $\eta$ and $Y_{\pi\pi}$. The window of $Y_{\pi\pi}$ is adjusted to Fig.\
 \protect\ref{ykpy}. }
 \label{hydsource}
\end{figure}

\begin{figure}[ht]
 \begin{center}
  \epsfig{file=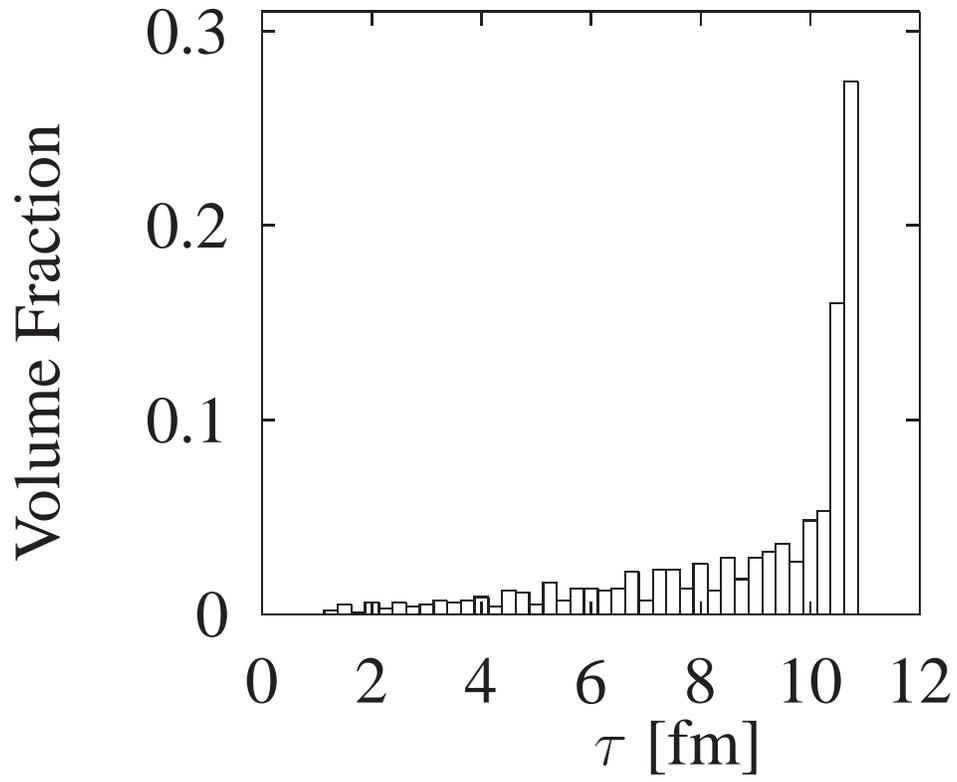}
 \end{center}
 \caption{The volume fraction of freeze-out hypersurface as a function of
 $\tau$.}
 \label{stau}
\end{figure}

\begin{figure}[ht]
 \begin{center}
  \epsfig{file=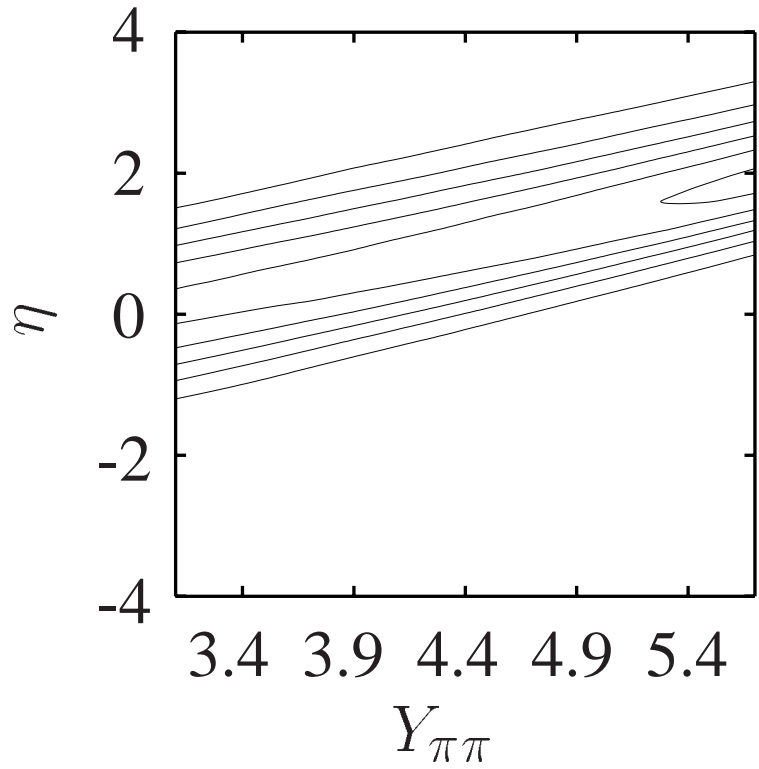}
 \end{center}
 \caption{Contour plot of source function for the Gaussian source model
 (\protect\ref{gsou}) as a function of $\eta$ and $Y_{\pi\pi}$.}
 \label{gsource}
\end{figure}

\begin{figure}[ht]
 \begin{center}
  \epsfig{file=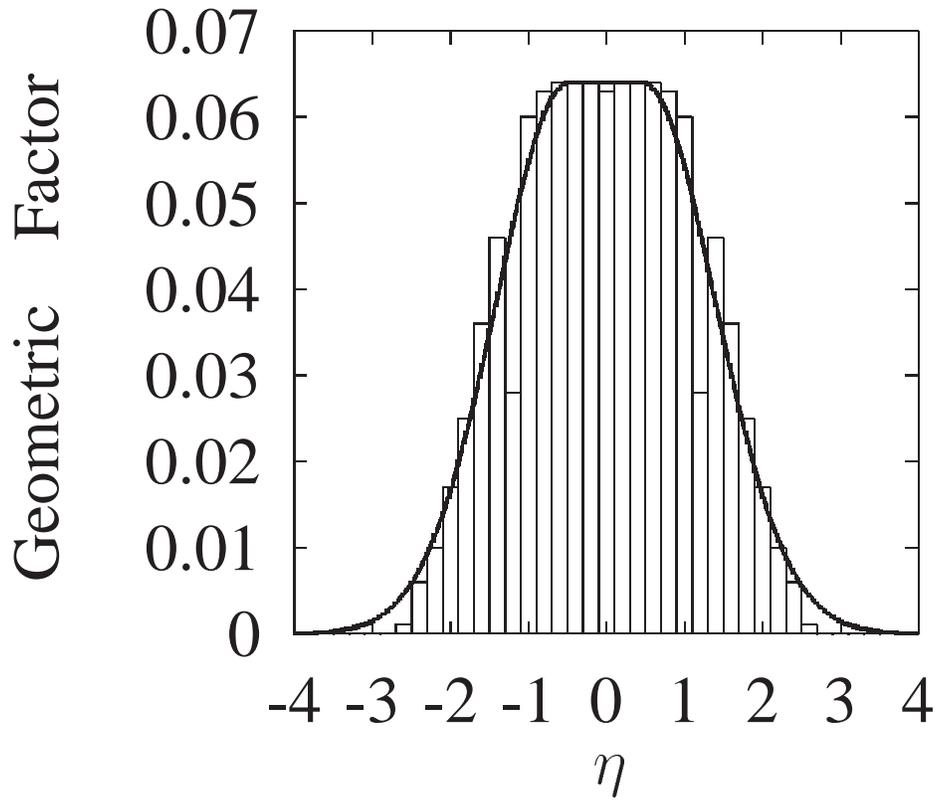}
 \end{center}
 \caption{The geometric factor of source function in $\eta$--space. The box
 profiles stand for our hydrodynamical model. The solid line stands for
 modified geometric factor $\tilde{G}(\eta)$ obtained from the fit in
 Eq.(\protect\ref{gflat}) where $\eta_0=0.92$ and $\delta \eta=0.47$ .}
 \label{seta}
\end{figure}

\begin{figure}[ht]
 \begin{center}
  \epsfig{file=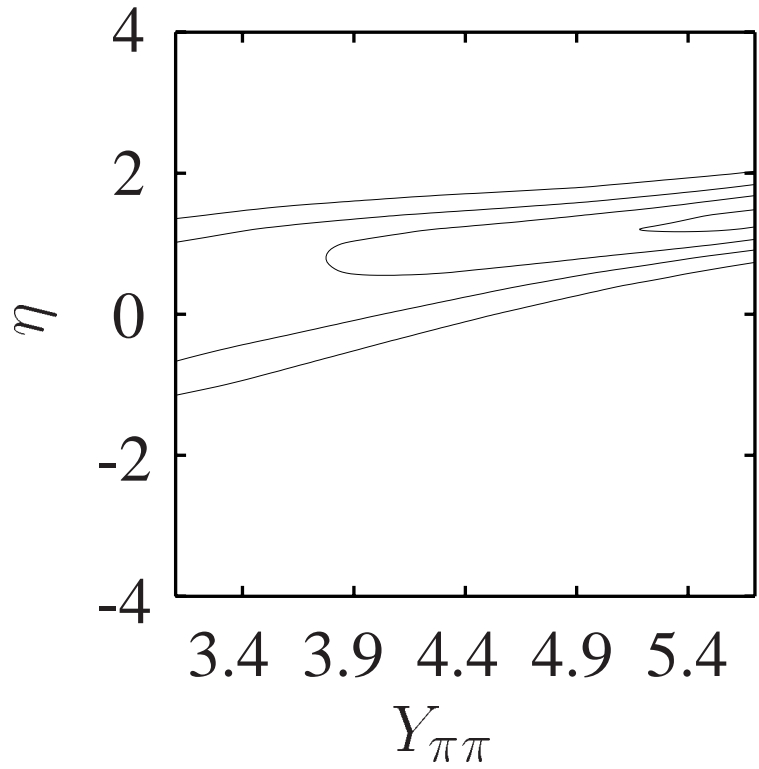}
 \end{center}
 \caption{Same as Fig.\ \protect\ref{gsource}, but $G(\eta)$ is now
 replaced by $\tilde{G}(\eta)$ in the calculation.}
 \label{gflatsource}
\end{figure}

In this section, we discuss the pair rapidity $Y_{\pi\pi}$ dependence
of YKP radii. The upper figure of Fig.\ \ref{ykpy} shows the $R_\bot$, which
is consistent with the experimental result. It is similar to the case of
$K_T$ dependence; our result of $R_0$ shown in the lowest figure becomes
smaller as $Y_{\pi\pi}$ increases and our $R_0$ does not agree with $\Delta
t$. This departure also can be explained as the result of source opacity. 
The middle figure stands for $R_\|$, which shows a similar tendency as the
experimental results. Source size $\Delta z$ is almost the same as the
space-time extension $R_\|$ (\ref{ykprl}). Hence the interpretation in the
YKP parametrization is applicable to our model. The disagreement of $R_\|$
with the experiment is due to smaller longitudinal source width at larger
$Y_{\pi\pi}$. In Fig.\ \ref{hydsource}, we can see that the width of the
source function $\Delta \eta=\sqrt{\langle \eta^2 \rangle-\langle \eta
\rangle^2}$ for $\eta$-direction becomes smaller as $Y_{\pi\pi}$
increases. As shown in Fig.\ref{gsource}, this property does not appear in a
Gaussian source model \cite{Chapman_HIP1} where the source function is given
as
\begin{equation}
 S_g(x^\mu,K^\mu)
  =\frac{m_T \cosh(\eta-Y_{\pi})}
  {(2\pi)^3\,\sqrt{2\pi (\delta \tau)^2}}
  \exp \left[-\frac{K^\mu U_\mu}{T}-\frac{(\tau-\tau_0)^2}{2(\delta \tau)^2}
                     -\frac{r^2}{2R^2}
                \right]G(\eta) \label{gsou},
\end{equation}
where
\begin{equation}
 G(\eta)=\exp\left[-\frac{(\eta-\eta_0)^2}{2(\delta \eta)^2}
\right].\label{etafac}
\end{equation}
Of course, a small $\Delta \eta$ does not always mean a small $\Delta z$ in
general,  because $z$ is given as $z=\tau \sinh\,\eta$.  
However, in the case where freeze-out takes place almost at some fixed $\tau$, 
$\Delta z$ is regarded as proportional to $\Delta \eta$. And this case 
can be applied to our model,  because for most of the particles, the
freeze-out occurs at the latest stage ($\tau \sim 10$ or 11) as shown in
Fig.\ \ref{stau}. The small width of the source function in our model is due
to a non-Gaussian feature of $\eta$-dependence of a geometric factor of the
source function, which corresponds to $G(\eta)$ in the Gaussian source
model. Figure\ \ref{seta} shows the corresponding normalized geometric
factor for the freeze-out hypersurface in our model; the geometric factor 
has a central plateau region. Non-Gaussian type source function comes from
both initial temperature distribution and longitudinal flow. A deformed
Boltzmann factor makes the source function narrower in a high $Y_{\pi\pi}$
region. This feature is not only in a hydrodynamical model but also appears in
thermal source models with a non-Gaussian geometric factor. In order to see
the effect clearly, we replace $G(\eta)$ in 
Eq.(\ref{gsou}) with the modified geometric factor 
$\tilde{G}(\eta)$.  The solid line in
Fig.\ref{seta} denotes the modified geometric factor $\tilde{G}(\eta)$ 
obtained by fitting the geometric factor for our model through  
\begin{equation}
 \tilde{G}(\eta)=
  \exp\left[-\frac{(|\eta|-\eta_0)^2}{2(\delta \eta)^2}
  \theta(|\eta|-\eta_0)\right].\label{gflat}
\end{equation}
Using the above $\tilde{G}(\eta)$ in Eq.(\ref{gsou}) instead of $G(\eta)$, 
the same source function as in Fig.\ \ref{gsource} is evaluated, and the
result is shown in Fig.\ \ref{gflatsource}. We can see that the width of the
source function becomes narrower at a high $Y_{\pi\pi}$ region than a low
$Y_{\pi\pi}$ region in the figure. Hence, we can say that the strong
$Y_{\pi\pi}$ dependence of $R_\|$ comes from the geometric structure of the
pion source in the rapidity space.

\begin{figure}[ht]
 \begin{center}
  \epsfig{file=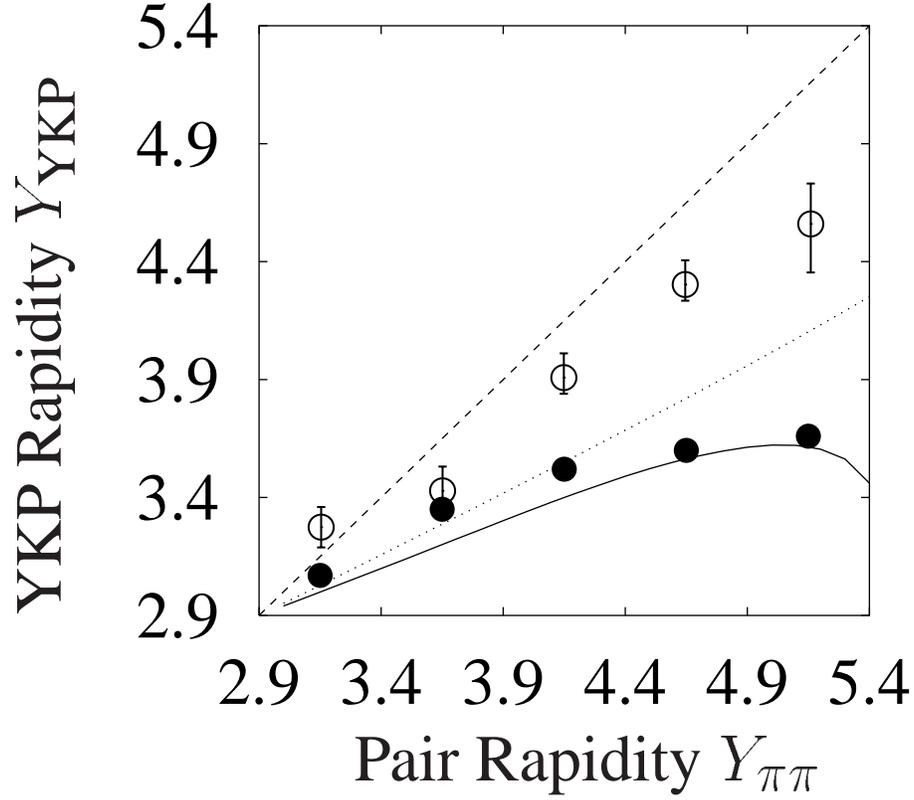}
 \end{center}
 \caption{$Y_{\pi\pi}$ dependence of $Y_{YKP}$. Open circles stand for the
 experimental results. Closed circles stand for our result obtained through
 Eq.\ (\protect\ref{yykp}). The solid line stands for our result
 calculated from space-time extensions Eq.(\protect\ref{ykpvelo}). The dotted
 line stands for average longitudinal expansion rapidity 
 $ \langle Y_L \rangle $ where $\langle \cdots \rangle $ is defined by
 Eq.(\protect\ref{average}). The dashed line stands for the infinite boost
 invariant source. }
\label{ykrap}
\end{figure}
\begin{figure}[ht]
 \begin{center}
  \epsfig{file=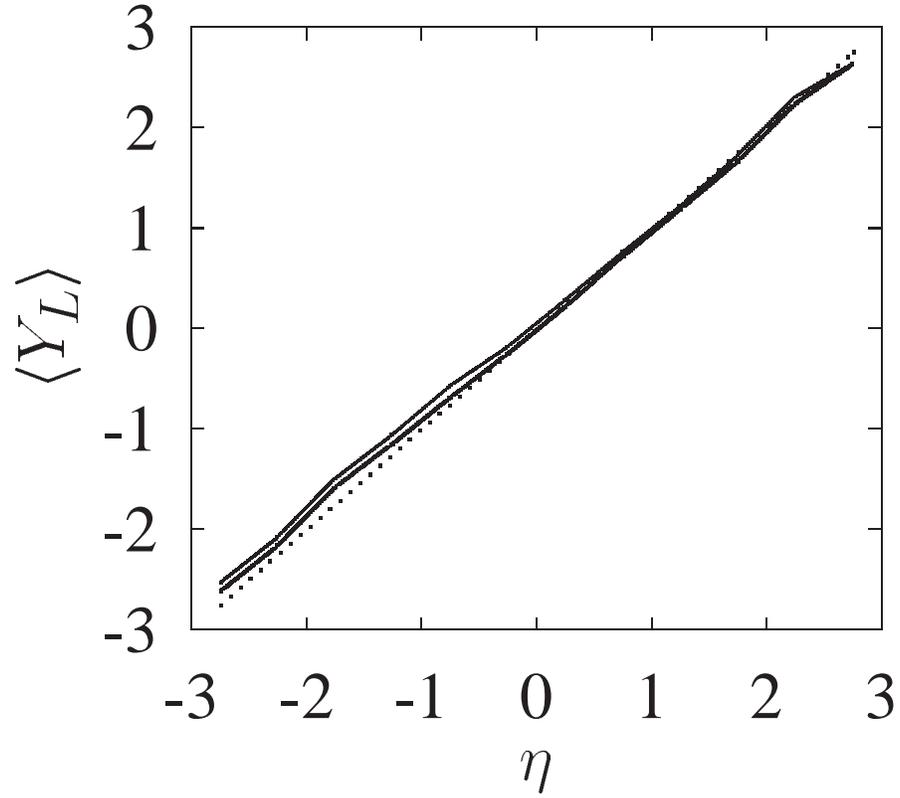}
 \end{center}
 \caption{Average longitudinal rapidity versus longitudinal coordinate
 $\eta$. The thick line stands for $Y_{\pi\pi}=4.9$, and the thin line stands
 for $Y_{\pi\pi}=3.4$. The dotted line shows Bjorken's scaling solution
 $Y_L=\eta$.}
 \label{scaling}
\end{figure}

Figure \ref{ykrap} shows pair rapidity $Y_{\pi\pi}$ dependence of the YKP
rapidity and the longitudinal expansion velocity. In our calculation, we used 
Eq.(\ref{ykp}) where $Y_{\text{YKP}}$ is derived by using the YKP 
velocity $v$, which is given as 
\begin{equation}
 Y_{\text{YKP}}=\frac{1}{2}\ln \frac{1+v}{1-v}+Y_{\text{cm}}.
 \label{yykp}
\end{equation}
The dashed line stands for the infinite boost-invariant source. The departure
between the solid line and the dotted line indicates that the failure of the
simple interpretation of YKP velocity, which might be caused opacity of the
source. The experimental results seem to be consistent with the infinite
boost-invariant source except for the high rapidity region. Our results are
much smaller than the boost-invariant model, as if our model contradicts the
boost-invariance. Figure\ \ref{scaling} shows average longitudinal rapidity
$\langle Y_L \rangle$ as a function of longitudinal coordinate $\eta$ for
slow ($Y_{\pi\pi}=3.4$) particles and fast ($Y_{\pi\pi}=4.9$) particles where
$K_T$ is fixed at 150 MeV. We can clearly see $\langle Y_L \rangle = \eta$
for both cases. Figure\ \ref{yl-eta} shows that $Y_L-\eta$ is positive for
small $\tau$ but is negative in a later stage of the space-time
evolution. Consequently, average rapidity $\langle Y_L \rangle$ is almost the 
same as $\eta$. Hence there must be another reason why $Y_{\text{YKP}}$ shows
non-boost-invariant behavior. By making use of the source model, the result
of the boost-invariant source is obtained by $\delta \eta \rightarrow \infty$
in Eq.(\ref{gsou}). For the finite but large enough $\delta \eta$, the peak of
the source function is only slightly shifted from $\eta=Y_{\pi\pi}$, which is
the peak of the Boltzmann factor $\exp(-m_T \cosh(\eta-Y_{\pi\pi})/T)$.
But for the small $\delta\eta$, a large shift will occur. Hence, the small
$Y_{\text{YKP}}$ reflects the finite size effect rather than longitudinal
expansion.

\section{Concluding Remarks}
\label{sec:Conclusion}

In this paper, with the use of the numerical simulation for the 
CERN-SPS 158 A GeV Pb+Pb collisions based on a relativistic
hydrodynamical model with first order phase transition, we investigated the
space-time structures via two-pion correlations in detail.  In addition to the
Cartesian parametrization, we adopted the Yano-Koonin-Podgoretski\u{\i}
parametrization for two-particle correlation function and compared $K_T$ and
$Y_{\pi\pi}$ dependence of source parameters with the experimental
result. In order to analyze the validity of the usual interpretations of HBT
radii, we also evaluated space-time extent and source size for freeze-out
hypersurface. 

In the case of Cartesian parametrization, we have compared $K_T$ dependence
of HBT radii and obtained results mostly consistent with the experiment. But
in the case of YKP parametrization, $K_T$ and $Y_{\pi\pi}$ dependence of the
transverse HBT radii, $R_\bot$, are consistent with the experiment. The $K_T$
dependence suggests the existence of the transverse flow, and the difference
between our calculation and the experiment can be improved by including
resonance decay into our model. The $K_T$ dependence of longitudinal HBT radii,
$R_\|$, agrees with the experiment but $Y_{\pi\pi}$ dependence does not, which
reflects a non-Gaussian structure of the geometric factor of the source
function in $\eta$--space. For these two HBT radii, usual interpretations of
the transverse and longitudinal source size are available.  
But in the case of temporal source radii $R_0$, the interpretation of the time
duration does not seem applicable at high $K_T$ and $Y_{\pi\pi}$. 
This comes from the fact that though first order phase transition exists,
freeze-out hyper surface in our simulation is an opaque source.
In the present stage, the experimental result is not clear whether the source
is opaque or not \cite{NA49_EPJC2,NA44_PRC58}, and we need more analysis based on realistic source models.

\section*{acknowledgement}
The authors are much indebted to Prof.\ I.\ Ohba and Prof.\ H.\ Nakazato for
their fruitful comments. They also would like to thank Dr.\ T.\ Hirano for
helpful discussions. This work was partially supported by a Grant-in-Aid for
Science Research, Ministry of Education, Science and Culture, Japan
(No.09740221) and Waseda University Media Network Center.

\appendix
\section*{Space-Time Evolution}
\begin{figure}[ht]
 \begin{center}
  \epsfig{file=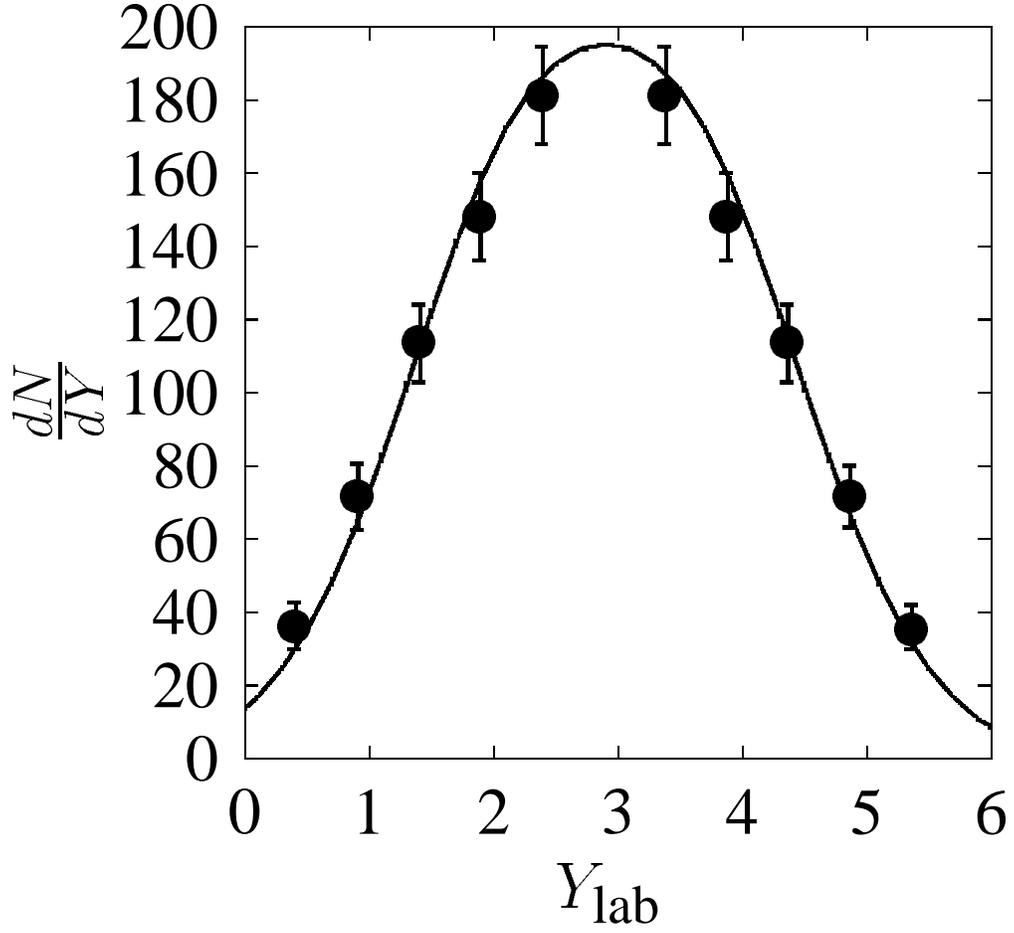}
 \end{center}
 \caption{Rapidity distribution of primary negative charged hadrons
 ($\pi^-, K^-, \overline{p}$). Closed circles are taken from NA49 data
 \protect\cite{NA491pd_QM96}. The solid line stand for our numerical result.}
 \label{dndy}
\end{figure}

\begin{figure}[ht]
 \begin{center}
  \epsfig{file=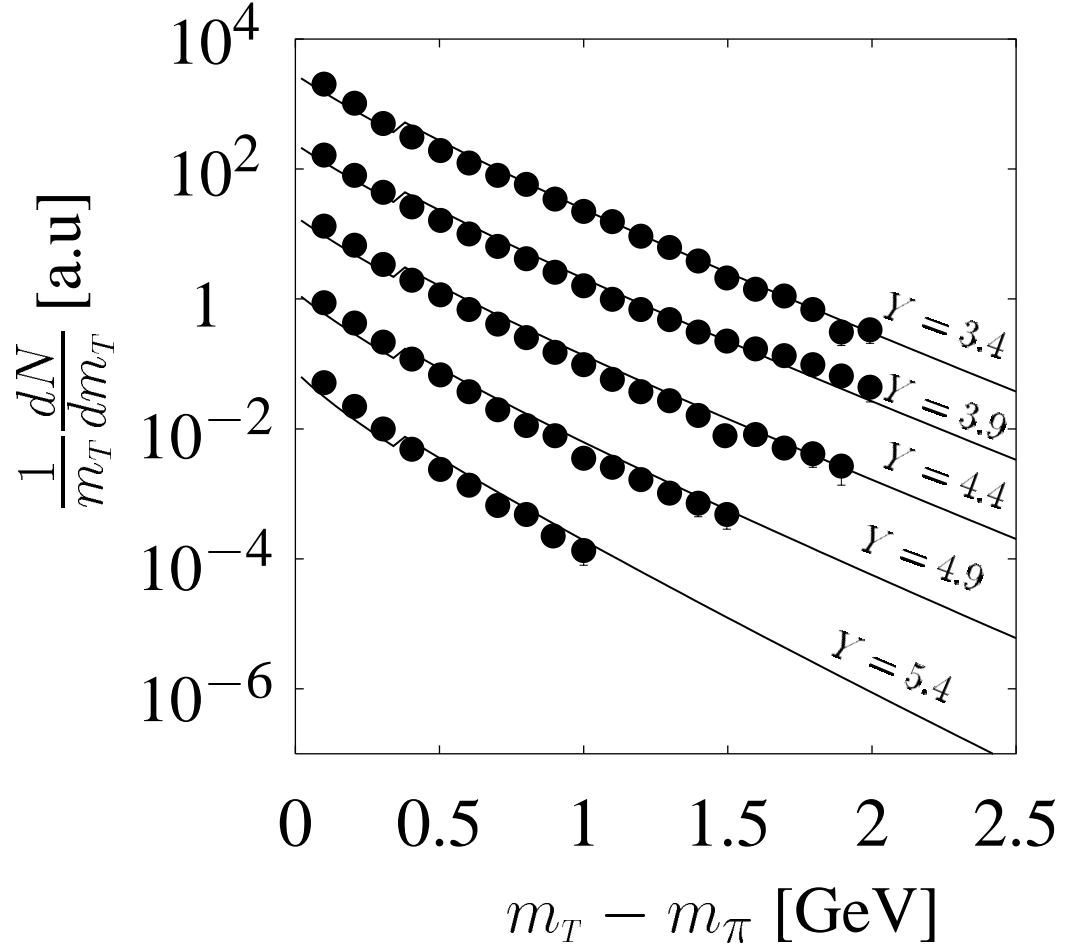}
 \end{center}
 \caption{Transverse mass distribution of negative charged hadrons
 ($\pi^-, K^-, \overline{p}$) successively scaled down by an order of
 magnitude in the rapidity range $y_\pi \in \{3.4,5.4\}$. Closed circles are
 taken from NA49 data \protect\cite{NA491pd_QM96}. The solid lines stands for
 our numerical results.}
 \label{mt}
\end{figure}

Assuming that local equilibrium is achieved at 1 fm later than the collision 
instance, we put initial conditions of the hydrodynamical model on 
the $\tau = \tau _{0}=$ 1 fm hypersurface.  
As for the initial local velocity, we use Bjorken's scaling solution
\cite{Bjorken_PRD27}, $Y_{\rm L} = \eta $, in the longitudinal direction 
 and neglect transverse flow, $Y_{\rm T} = 0$.  
Initial entropy density distribution is given as,
\begin{equation}
S(\tau_0,\eta)=S(T_0)\exp\left[-{\frac{(\vert \eta \vert-\eta_0)^2}
{2 \cdot{\sigma_\eta}^2}} \theta(\vert \eta \vert-\eta_0)
-{\frac{(r-r_0)^2}{2 \cdot{\sigma_r}^2}} \theta(r-r_0)\right]
\label{(3)},
\end{equation}
and parameters we used in this paper are summarized in Table
\ref{initialparam} .
Those values are so designed to reproduce hadronic distributions of
the recent experimental results with freeze-out temperature as $T_{\rm
f}=140$ MeV. Figures \ref{dndy} and \ref{mt} show the rapidity distribution
and the transverse mass distribution of negative charged hadrons in Pb+Pb 158
A GeV collisions by NA49 \cite{NA491pd_QM96}. 

\begin{figure}[ht]
 \begin{center}
  \epsfig{file=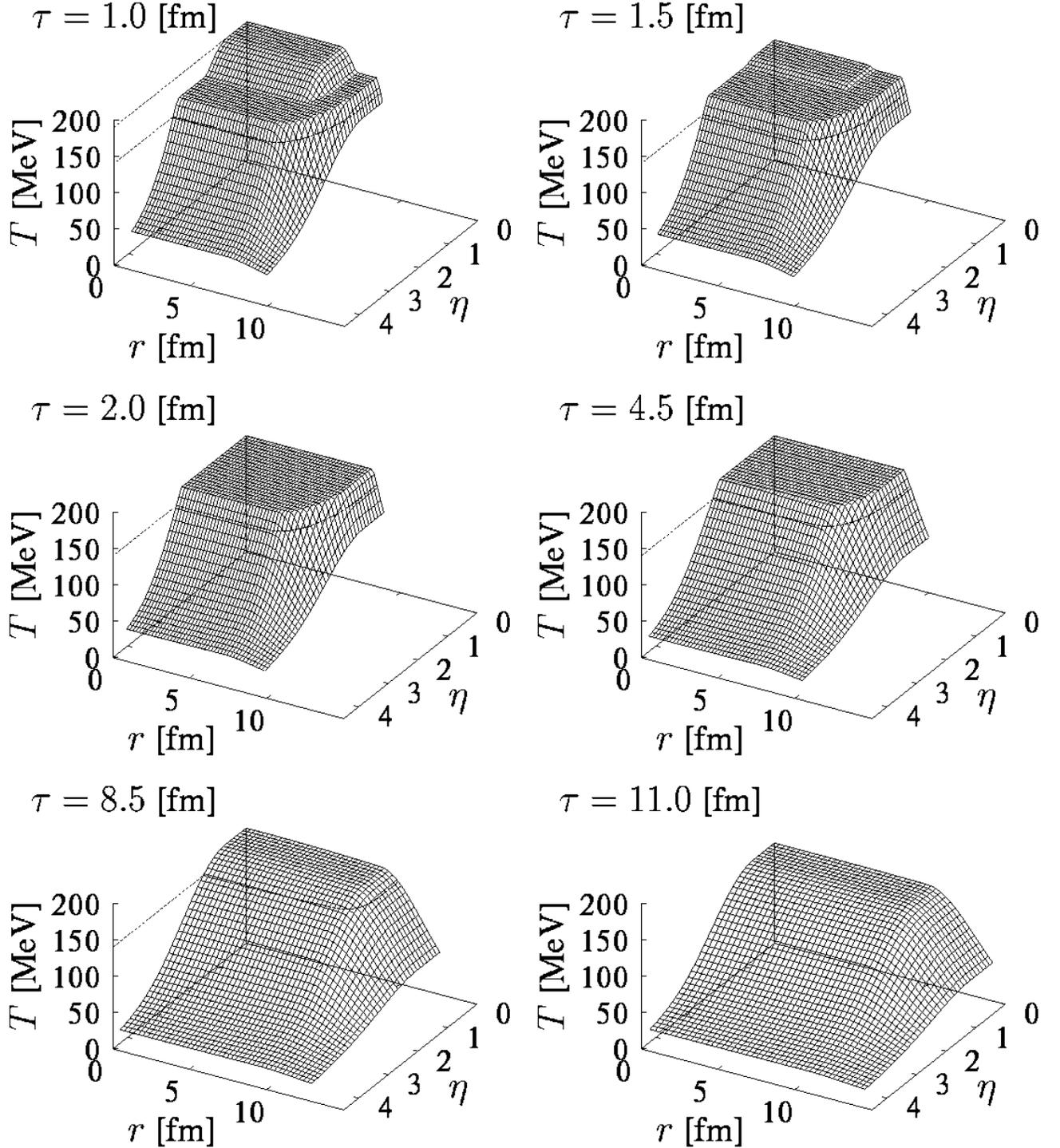}
 \end{center}
 \caption{The space-time evolution of temperature-distribution with initial
 temperature (190 [MeV]), critical temperature (160 [MeV]) and freeze-out
 temperature (140 [MeV]). The flat area corresponds to the mixed phase
 region. From these graphs we can see that the QGP phase vanishes at $\tau$ =
 2.0 [fm] and the mixed phase vanishes at $\tau$ = 8.5 [fm], and the
 temperature becomes less than the freeze-out temperature everywhere in the
 fluid at $\tau$ = 11.0 [fm]. }
 \label{temperature}
\end{figure}

Figure \ref{temperature} displays the space time evolution of the 
temperature distribution of our hydrodynamical 
model with the above parameters.
From this simulation we can estimate ``lifetimes'' of fluid, QGP phase 
and mixed phase which are summarized in Table \ref{lifetime}.

In our simulation, the isothermal region at phase transition temperature 
$T=T_{\rm c}$ appears very clearly. Though the global structure of 
the chronological 
evolution does not differ essentially from the model with smooth phase 
transition in a small temperature region \cite{Akase_PTP85,Sollfrank_PRC55},
  by virtue of the use of the volume fraction $\lambda$ at phase 
transition temperature, we can easily pick up the volume element with 
just the phase transition temperature.

\begin{figure}[ht]
 \begin{center}
  \epsfig{file=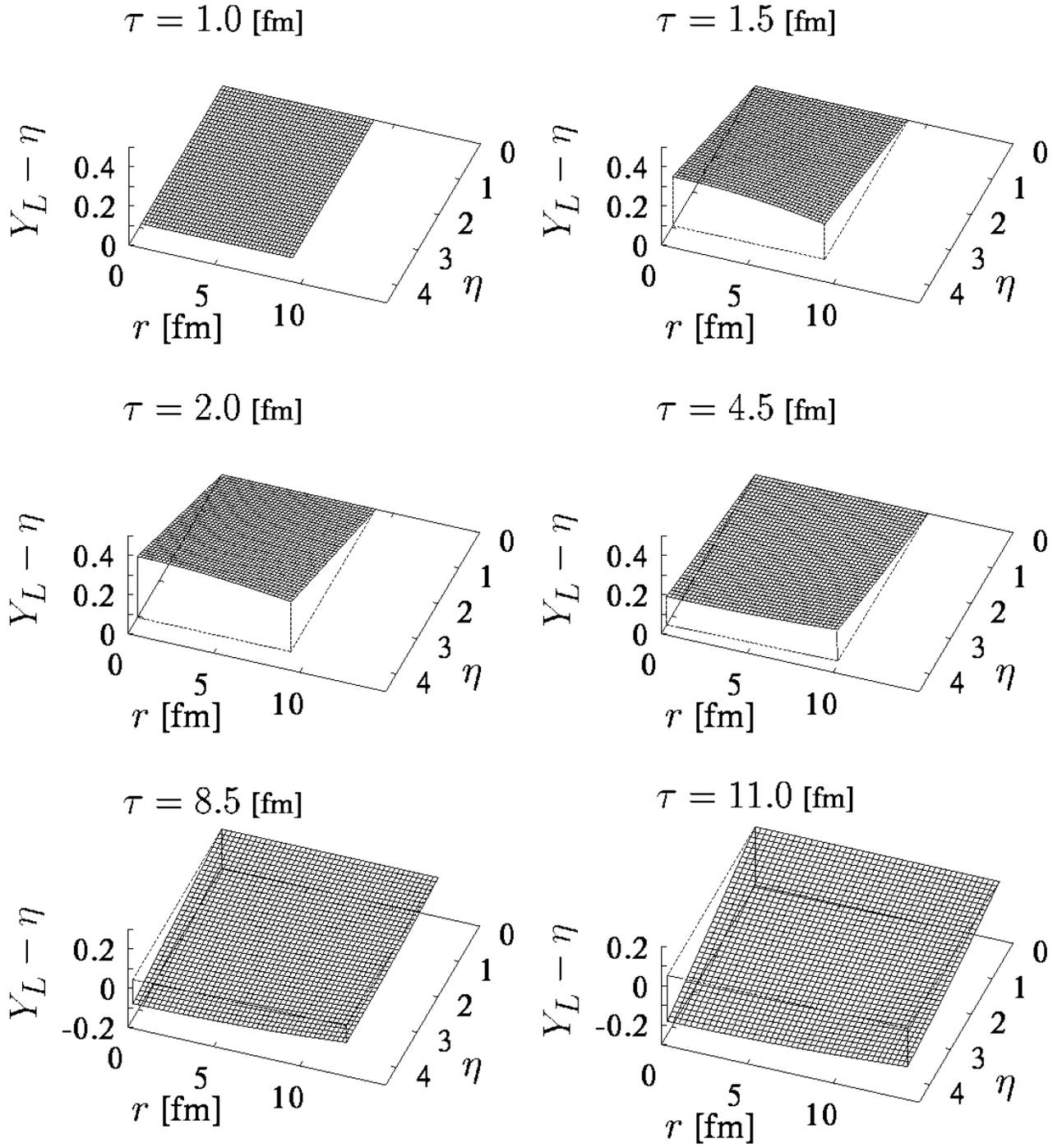}
 \end{center}
 \caption{The space-time evolution of $Y_{L} - \eta$.}
 \label{yl-eta}
\end{figure}

\begin{figure}[ht]
 \begin{center}
  \epsfig{file=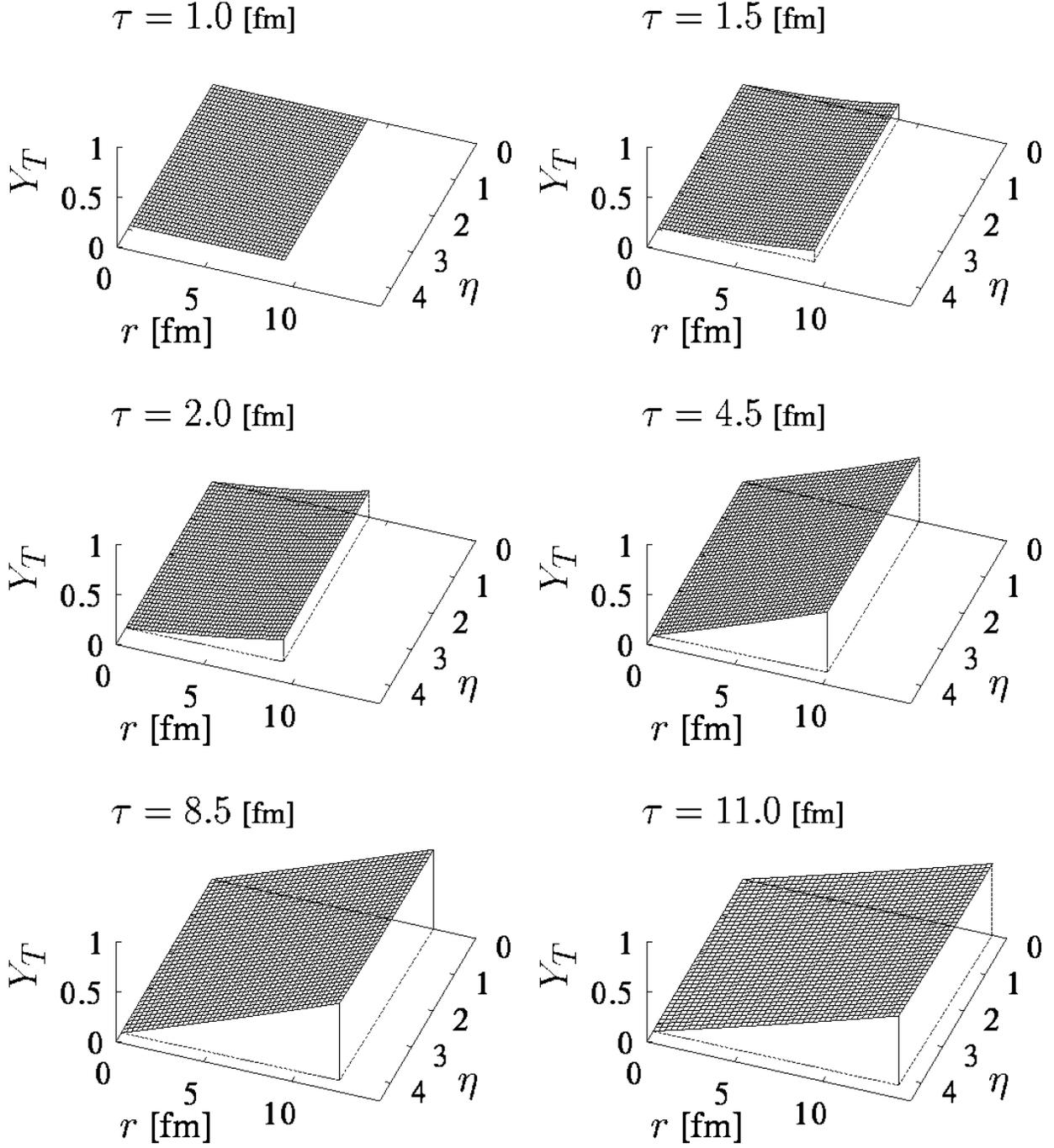}
 \end{center}
 \caption{The space-time evolution of $Y_{T}$.}
 \label{yt}
\end{figure}

We describe the space time evolution of the local velocity in terms of
$Y_{L}-\eta$, $Y_{T}$. Figure \ref{yl-eta} shows that  
$Y_{L}-\eta$ becomes negative after $\tau=8.5$ fm and  
the difference between our numerical solution and Bjorken's 
scaling solution is not significant on the whole. 
Figure \ref{yt} indicates that the dependence of $Y_{T}$ on $\eta$ is
small and  $Y_{T}$ is proportional to $r$.

\begin{figure}[ht]
 \begin{center}
  \epsfig{file=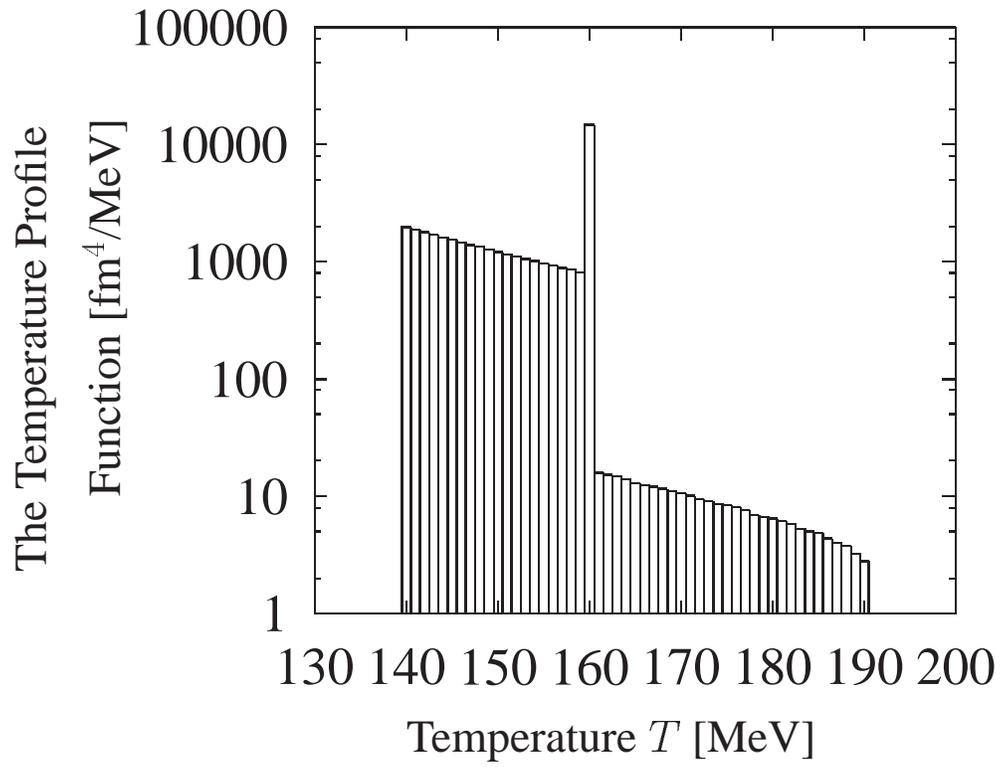}
 \end{center}
 \caption{The temperature profile function.}
 \label{profile}
\end{figure}

The temperature profile function (Fig.\ \ref{profile}) shows the significant
contribution of the  space-time volume at $T=T_{\rm c}$.

\newpage
\begin{table}[ht]
 \caption{Initial Parameters}
 \begin{tabular}{cccccc}
  & $T_{0}$ (MeV) & $\eta_{0}$ & $\sigma_{\eta}$ & $r_{0}$ (fm) & 
 $\sigma_{\rm r}$ (fm) \\ \hline 
 Pb + Pb& 190 & 0.7 & 0.7 & $1.2 \times (207)^{(1/3)}-1.0 $ & 1.0 \\
 \end{tabular}
 \label{initialparam}
\end{table}

\begin{table}[ht]
 \caption{Outputs}
 \begin{tabular}{lc}
 Maximum initial energy density &  3.1 GeV/fm$^{3}$ \\ \tableline
 Maximum lifetime of fluid & 11.0 fm  \\ \tableline
 Maximum lifetime of QGP phase  & 2.0 fm \\ \tableline
 Maximum lifetime of mixed phase & 8.5 fm\\ 
 \end{tabular}
 \label{lifetime}
\end{table}

\newpage

\newpage

\newpage

\newpage

\newpage

\newpage

\newpage

\newpage

\newpage

\newpage

\newpage

\newpage

\newpage

\newpage

\newpage

\newpage

\end{document}